\documentclass[twocolumn]{aastex62}

\usepackage{graphicx}
\usepackage{dcolumn}
\usepackage{bm}
\usepackage{color}
\usepackage{xcolor}
\usepackage{amsmath}
\usepackage{lineno}

\newcommand\lsim{\mathrel{\rlap{\lower4pt\hbox{\hskip1pt$\sim$}}
\raise1pt\hbox{$<$}}}
\newcommand\gsim{\mathrel{\rlap{\lower4pt\hbox{\hskip1pt$\sim$}}
\raise1pt\hbox{$>$}}}
\graphicspath{{./}{figures/}}

\revised{\today}
\submitjournal{ApJ}

\begin{document}

\title{Detecting Gravitational Wave Bursts From Stellar-Mass Binaries in the Milli-hertz Band}

\correspondingauthor{Zeyuan Xuan}
\email{zeyuan.xuan@physics.ucla.edu}

\author{Zeyuan Xuan}
\affiliation{ Department of Physics and Astronomy, UCLA, Los Angeles, CA 90095}
\affiliation{Mani L. Bhaumik Institute for Theoretical Physics, Department of Physics and Astronomy, UCLA, Los Angeles, CA 90095, USA}

\author{Smadar Naoz}
\affiliation{ Department of Physics and Astronomy, UCLA, Los Angeles, CA 90095}
\affiliation{Mani L. Bhaumik Institute for Theoretical Physics, Department of Physics and Astronomy, UCLA, Los Angeles, CA 90095, USA}

\author{Bence Kocsis}
\affiliation{Rudolf Peierls Centre for Theoretical Physics, Parks Road, Oxford OX1 3PU, UK}

\author{Erez Michaely}
\affiliation{ Department of Physics and Astronomy, UCLA, Los Angeles, CA 90095}
\affiliation{Mani L. Bhaumik Institute for Theoretical Physics, Department of Physics and Astronomy, UCLA, Los Angeles, CA 90095, USA}

\begin{abstract}
{The dynamical formation channels of gravitational wave (GW) sources typically involve a stage when the compact object binary source interacts with the environment, which may excite its eccentricity, yielding efficient GW emission. For the wide eccentric compact object binaries, the GW emission happens mostly near the pericenter passage, creating a unique, burst-like signature in the waveform. This work examines the possibility of stellar-mass bursting sources in the mHz band for future LISA detections. Because of their long lifetime ($\sim 10^{7}\rm\, yr$) and promising detectability, the number of mHz bursting sources can be large in the local universe. For example, based on our estimates, there will be $\sim 3 - 45$ bursting binary black holes in the Milky Way, with $\sim 10^{2} - 10^{4}$ bursts detected during the LISA mission. Moreover, we find that the number of bursting sources strongly depends on their formation history. If certain regions undergo active formation of compact object binaries in the recent few million years, there will be a significantly higher bursting source fraction. Thus, the detection of mHz GW bursts not only serves as a clue for distinguishing different formation channels, but also helps us understand the star formation history in different regions of the Milky Way. 
}
\end{abstract}

\keywords{gravitational waves -- detection}

\section{Introduction} \label{sec:intro}
Gravitational wave bursts are expected to be a natural consequence of wide, eccentric compact object binaries' GW emission. In particular, for highly eccentric binaries, the GW emission happens mostly near the pericenter passage, creating a unique, pulse-like signature that lasts much shorter than the orbital period \citep[see, e.g.,][]{Kocsis+2006,O'Leary+09,Gould_2011,Kocsis_2012,Seto_2013}. Most of the burst emission does not yield a merger immediately; therefore, if the GW sources have an orbital period that is smaller than the observational time, the bursting signal will appear as {\it repeated bursts} (RB).

As the source of GW bursts, eccentric compact object binaries are often expected to form via dynamical channels, during which the binary's interaction with the environment excites its eccentricity $e$, reduces the pericenter distance, and causes effective GW radiation.
Generally, several dynamical mechanisms involve a highly eccentric stage during the GW source's evolution. For example, in a hierarchical triple system (a tight binary orbiting a third body on a much wider “outer orbit"), the inner binary can undergo large eccentricity oscillations due to gravitational perturbations from the tertiary. This so-called eccentric Kozai-Lidov (EKL) mechanism \citep{Kozai1962,Lidov1962,Naoz16} may potentially contribute to the overall merger rate of stellar mass compact objects at significant levels \citep[e.g.,][]{wen03,Hoang+18,Hamers+18,Stephan+19,Bub+20,Deme+20,Wang+21}. Additionally, 
wide compact object binaries in the galactic field may interact with the surrounding environment, which can excite the binary's eccentricity \citep[e.g.,][]{Michaely+19,Michaely+20,Michaely+22}, resulting in merger events that are potentially observable. Moreover, a variety of dynamical mechanisms, such as GW capture, binary-single, and binary-binary scattering interaction, can take place in dense star clusters, producing compact object binaries with non-negligible eccentricity \citep[e.g.,][]{O'Leary+09,Thompson+11,Aarseth+12,Kocsis_2012,breivik16,Orazio+18,Zevin_2019,Samsing+19,Martinez+20,Antonini+19,Kremer_2020,wintergranic2023binary}. Furthermore, dynamical interactions in a flattened black hole distribution, such as in a stellar disk or an active galactic nucleus accretion disk, may also lead to highly eccentric mergers \citep{Tagawa+2021,Samsing+2022,Munoz+22,Gautham+23}.

The detection of repeated bursts can greatly enhance our understanding of the GW sources' dynamical formation. In particular, with the corresponding data analysis methods \citep[e.g.,][]{Tai_2014,Loutrel_2017,B_csy_2020}, the LIGO-VIRGO-KAGRA (LVK) detectors may detect the residual eccentricity of highly eccentric sources, thus helping with identifying the fraction of GW sources formed in a variety of dynamical channels \citep[e.g.,][]{east13,samsing14,Coughlin_2015,Gondan_2018a,Gondan_2018b,Zevin_2021}. Furthermore, as one of the main targets of the future space-based gravitational wave detectors, the extreme mass ratio inspirals (EMRIs) are expected to have high eccentricity in the mHz band, emitting detectable GW bursts at a cosmological distance
\citep[e.g.,][]{Glampedakis_2005,Hopman2006,Rubbo06,Amaro_Seoane_2007,Barack_2009,Berry_2013,Chen_2018,Fan22,oliver2023gravitational,Naoz+22,Naoz+23}. With the unique signature of highly-eccentric orbit, we can even enhance the peculiar acceleration measurement of the GW source by a factor of $\sim 100$ \citep{Xuan23} and probe the gravitational potential surrounding these bursting systems \citep{Zhang_2021,Romero23}.

In this work, we focus on stellar-mass compact object binaries that are bursting in the mHz band. These sources will typically have a semi-major axis $a\gsim 0.1 $~au, up to $\sim 10^{5} $~au, with the pericenter distance $ a(1-e)\sim 10^{-3} $~au, which makes the peak frequency of GW emission in the milli-hertz band. Although the GWs from these sources may be weaker because of their light mass (compared with EMRIs) and larger orbital separation \citep[compared with bursting sources in the LIGO band, see, e.g., ][]{Randall_2022}, they are expected to have a much longer lifetime and therefore a larger number of systems in the mHz band \citep[e.g.,][]{Fang_2019}. Furthermore, since the mHz bursting sources
have not undergone significant orbital shrinkage, their eccentricity is more likely to be extreme, and their evolution is less rapid. These features can be used to extract relevant information about compact object binaries' surrounding environment, as well as their formation history.

This paper is organized as follows. We first specify the definition of bursting GW sources in Section~\ref{sec:definition}, then quantify their detectability (Section~\ref{subsec:detectability}) and lifetime (Section~\ref{subsec:lifetime}). In Section~\ref{sec:population}, we constrain the population of bursting sources from the observational results of LIGO (Section~\ref{sec:ligoconstrain}), then carry out numerical simulations to predict the number of detectable bursts for LISA. We show the results separately, for the field (Section~\ref{sec:field}), globular clusters (Section~\ref{sec:gcs}), and the galactic nucleus (Section~\ref{sec:GN}) of the Milky Way. In Section~\ref{sec:discussion}, the properties of stellar-mass bursting sources are summarized and the implications are discussed.

Unless otherwise specified, we set $G=c=1$.

\section{Repeated bursts for eccentric sources in the milli-hertz band}
\label{sec:RB property}
\subsection{Burst Definition}
\label{sec:definition}
Heuristically, a burst is defined as a situation in which a significant amount of energy is emitted in a short amount of time compared to the orbital period. In particular, for eccentric binary sources, the pericenter time is usually defined with the orbital separation, velocity, and eccentricity, $(r_p,v_p,e)$ as \citep[e.g.,][]{O'Leary+09}
\begin{equation}
   T_{\rm p} \sim \frac{r_p}{v_p} \sim (1-e)^{3/2} T_{\rm orb} \ ,
    \label{eq:time0}
\end{equation}
where $T_{\rm orb}=2\pi a^{3/2} m_{\rm bin}^{-1/2}$ is the period of a binary with a mass $m_{\rm bin}$ and semi-major axis $a$. Note that there is an order unity factor of $(1+e)^{-1/2}$ that we omit here. 

As a proof of concept, consider a source with  $e>0.9$, 
we demonstrate below that, in this case, more than $88\%$ of the GW energy emission is emitted in less than $1\%$ of the orbital period during pericenter passage, regardless of its masses and semi-major axis.

The fraction of GW energy emitted during a burst can be estimated by integrating the power of GW emission, $P$, over the orbital true anomaly $\psi$. Particularly, $P(\psi)$ is given by \citep{peters63}:
\begin{eqnarray}
P(\psi) & = & \frac{8}{15}  \frac{m_1^2 m_2^2\left(m_1+m_2\right)}{a^5\left(1-e^2\right)^5}(1+e \cos \psi)^4 \nonumber \\ 
& \quad & \times \left[12(1+e \cos \psi)^2+e^2 \sin ^2 \psi\right] \ ,
\label{eq:ppsi}
\end{eqnarray}
in which $m_1, m_2$ are the mass of the binary's components.

Throughout the paper, as a proof of concept, we adopt $1\%$ as the fraction of burst time relative to the orbital period (note that other bursting fractions are straightforward to analyze.) In this case, the corresponding change in the orbital phase, $\delta \psi$, can be solved by integrating the Keplerian motion of the binary near the pericenter passage, $\psi\sim 0$, and require that $T_{\rm burst}/T_{\rm orb}=1\%$:
\begin{equation}\label{eq:Tfrac}
 \frac{T_{\rm burst}}{T_{\rm orb}} = \frac{1}{T_{\rm orb}} \int_{-\frac12\delta\psi}^{\frac12\delta\psi} \frac{d\psi}{\dot\psi} = \frac{1}{2\pi} \int_{-\frac12\delta\psi}^{\frac12\delta\psi} \frac{(1-e^{2})^{\frac{3}{2}}}{(1+e\cos\psi)^2} {d\psi} \ ,
\end{equation}
where $\dot{\psi}$ is the time derivative of the orbital true anomaly:
\begin{equation}
\dot{\psi}=\frac{\sqrt{\left(m_1+m_2\right) a\left(1-e^2\right)}}{d^2} \ ,
\end{equation}
with $d=a\left(1-e^2\right)/(1+e \cos \psi)$. 

The fraction of energy emitted in the burst, compared with the total energy loss during one orbital period, is given by:
\begin{equation}
    \mathcal{F}(e)=\frac{1}{\langle P\rangle T_{\rm orb}}\int_{-\delta \psi/2}^{\delta \psi/2}P(\psi)  \frac{d\psi }{\dot{\psi}} \ ,
    \label{eq:fract}
\end{equation}
where $\langle P\rangle$ is the orbit-averaged GW energy emission power, given by \citet{peters63}:
\begin{equation}
\langle P\rangle=\frac{32}{5}  \frac{m_1^2 m_2^2\left(m_1+m_2\right)}{a^5\left(1-e^2\right)^{7 / 2}}\left(1+\frac{73}{24} e^2+\frac{37}{96} e^4\right) \ .
\end{equation}

\begin{figure}[htbp]
    \centering
    \includegraphics[width=\linewidth]{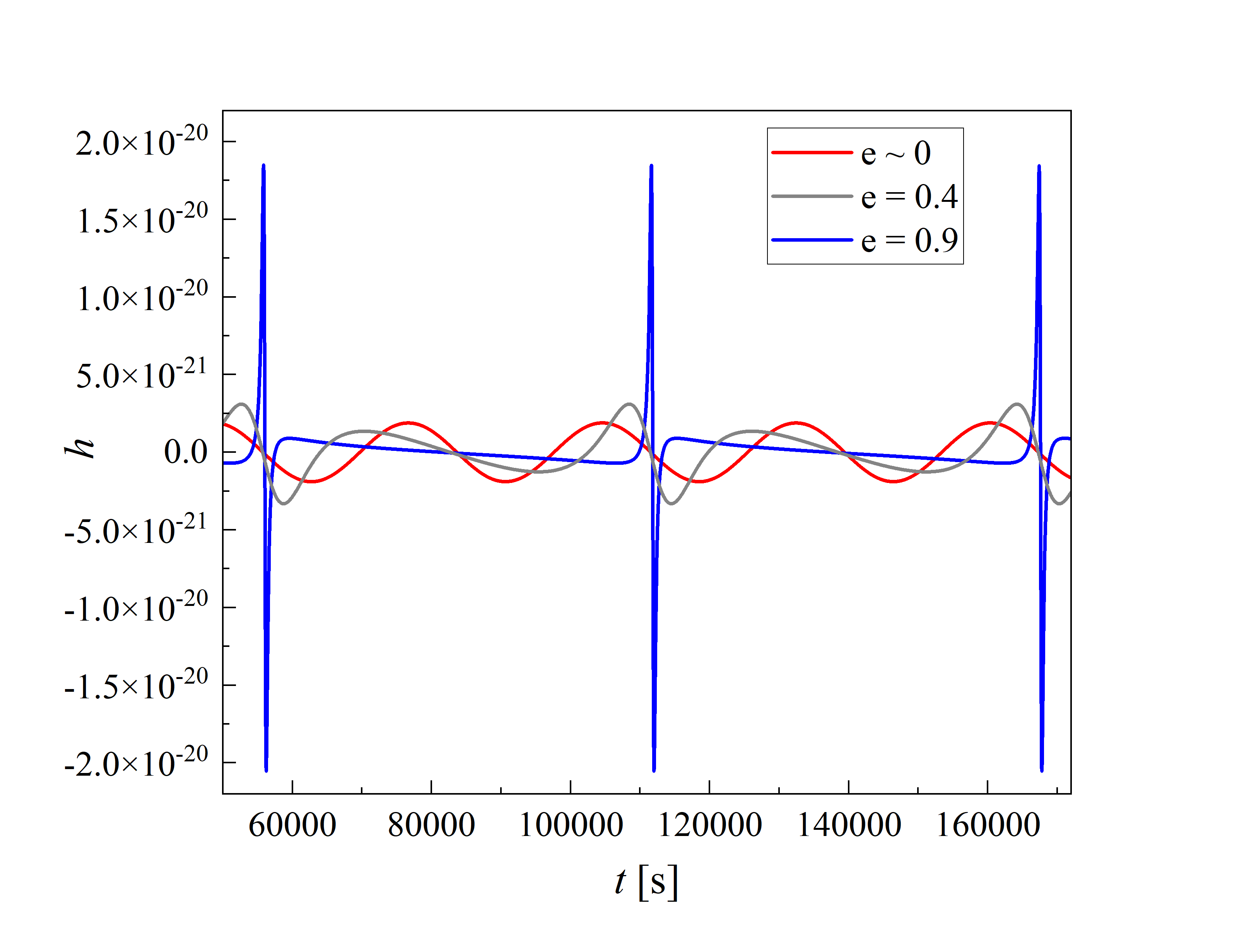}
    \caption{{\bf{The GW waveforms of a BBH system with the same orbital frequency but different eccentricities.}} We show a BBH system with  $m_{1}=m_{2}=20$~M$_{\odot},$ 
    orbital frequency $f_{\rm orb}=1.8\times 10^{-5}$~Hz, luminosity distance
    $D_{l}=8$~kpc, and $e=0.001, 0.4, 0.9$, respectively. As explained in the text, when the binary's orbital eccentricity increases, its GW energy emission will concentrate near each pericenter passage, turning the GW signal from a sinusoidal wave ($e\sim 0$ case) into a ``burst-like" waveform ($e=0.9$ case).}
    \label{fig:diffe}
\end{figure}

We note that the fraction shown in Equation~(\ref{eq:fract}) is a function of the orbital eccentricity, as $e$ will affect both the GW emission waveform (Eq.~\ref{eq:ppsi}) and the Keplerian motion, resulting in different $\delta \psi$ (Eq.~\ref{eq:Tfrac}). Therefore, higher eccentricity values will yield a higher fraction of energy loss around the pericenter passage, making GW emission look more and more like bursts (see, e.g., Figure~\ref{fig:diffe}. The waveform in this case is calculated numerically using the x-model \citep[see, e.g., eq.(1)-(13), (A1)-(A36) in][]{Hinder+10}). 

Adopting $\mathcal{F}(e)\sim 90\%$ and $T_{\rm burst}/T_{\rm orb}=1\%$, we can solve for the corresponding eccentricity threshold and find that a bursting source should have $e>0.9$, regardless of its masses and semi-major axis. 

\subsection{Detectability of Bursting Sources in the Milli-hertz Band}
\label{subsec:detectability}
The detection of GW bursts from highly eccentric sources is quite different from that of quasi-circular sources. In particular, when the binary's eccentricity is small (e.g., $e\sim 0.001$ in Figure~\ref{fig:diffe}), the GW signal can be well approximated by a near-monochromatic, sinusoidal wave. Therefore, the GW templates of these sources are relatively straightforward to construct, which enables us to adopt the matched filtering method in the template fitting \citep{thorne1987300, Finn92,Cutler+94} and measure the source's parameters accurately.

On the other hand, when the source's eccentricity increases (e.g., $e=0.4, e=0.9$ in Figure~\ref{fig:diffe}), its GW emission will become stronger upon each pericenter passage, turning the signal into a burst-like waveform. For example, mHz bursting sources typically emit a bright GW signal during a period of $10^{2}\sim 10^{3}$~s. However, for the rest of the orbital time (which is typically much longer than the signal time and can be days, months, or even years), the emission is suppressed.  The transient nature of bursting sources makes them hard to detect, mostly because it adds to the computational expense of the template fitting and reduces their average signal-to-noise ratio.

Note that the transient nature of bursting sources can also lead to their long lifetime. Since the bursting sources spend a small amount of time emitting GW during each orbit (near the pericenter passage), the average power of orbital energy loss is small when compared with a circular binary with the same GW frequency as the eccentric bursting source's peak frequency. This means their orbit will shrink on a longer timescale. Therefore, bursting sources potentially have more significant numbers in future mHz GW detection. In the following sections, we will quantify how this feature enhances their detectability. 

The signal-to-noise ratio (SNR) of a bursting source can be estimated analytically\footnote{Here we focus on giving a simple and direct estimation by comparing the system to a circular orbit with the radius of the pericenter distance. A more detailed analysis of the signal-to-noise ratio may be found in e.g. \citet{Kocsis2006,O'Leary+09,Randall_2022}.}. In particular, the energy loss of a bursting source mostly takes place near the pericenter passage; thus, consider a circular binary with the orbital radius the same as the bursting source's pericenter distance, $r_c=a(1-e)$, then such a circular binary's GW emission power should be on the same order of magnitude as the ``bursting power" of the bursting source.  

This approximation is consistent with the nature of GW radiation, since the strain amplitude is proportional to the second-order time derivative of the binary's mass-quadrupole moment, while the peak GW frequency depends on the
angular velocity at pericenter passage.
Thus, a circular and eccentric binary with a similar radius/periapsis and velocity 
have a similar GW strain amplitude and frequency. 

The bursting source has similar GW amplitude and frequency as the corresponding circular binary enclosed by the pericenter distance. The effective GW burst duration can be written as:
\begin{equation}
    T_{\rm burst} \sim T_p \sim T_{\rm orb} (1-e)^{\frac{3}{2}} \ ,
    \label{eq:time}
\end{equation}
Note that this expression also describes the period of the corresponding circular orbit.

Moreover, highly eccentric sources emit GW emission in a wide range of frequencies, which peak at $f_{\rm peak} \sim f_{\rm orb} (1-e)^{-3/2}$ \citep{peters63}. In the approximation that the GW burst emission from a highly-eccentric source has a similar frequency as the circular orbit with $r_c\sim a(1-e)$, the peak frequency is approximately\footnote{The peak frequency is often estimated as $f_{\rm peak} = f_{\rm orb} (1+e)^{1/2} (1-e)^{-3/2}$ \citep{O'Leary+09}. For consistency with other definitions in our treatment above, we adopt $f_{\rm circ, GW}$, which differs by an order of unity.}:
\begin{equation}
    f_{\rm burst}\sim f_{\rm circ, GW}=\frac{2}{T_{\rm circ}(r_c)} \ ,
    \label{eq:frequency}
\end{equation}
in which $f_{\rm circ, GW}$ is the GW frequency of the corresponding circular orbit, and $T_{\rm circ}$ is the period of the circular orbit.

Similarly, the strain amplitude of a burst can be estimated by considering the circular orbit enclosed within the pericenter distance \citep{peters63,Kocsis_2012}, i.e., 
\begin{align}
    &h_{\rm burst} \sim \sqrt{\frac{32}{5}}\frac{m_1m_2}{D_l r_c}\sim \sqrt{\frac{32}{5}}\frac{m_1m_2}{D_la(1-e)}\nonumber\\
    &\sim  5.54\times 10^{-21}\eta_s \left(\frac{m}{10\,\rm M_{\odot}}\right)^{\frac{5}{3}} \left(\frac{T_{\rm burst}}{1000\,\rm s}\right)^{-\frac{2}{3}}\left(\frac{D_{l}}{8\,\rm kpc}\right)^{-1}\ ,
    \label{eq:amplitude}
\end{align}
here $D_{l}$ is the luminosity distance of the binary, and the constant coefficient comes from the average of strain amplitude over the binary's orientation. Furthermore, $m=(m_1+m_2)/2$ is the average component mass, $\eta_s=4 m_1m_2/(m_1+m_2)^{2}= 4 q/(1+q)^2$ is unity for equal mass sources where $q=m_1/m_2$.

The analytical expression of signal-to-noise ratio (SNR), for a monochromatic source, is given by \citep[see, e.g.,][]{Seto02WD}: 
\begin{equation}
{\rm{S N R}}=\frac{h_A \sqrt{fT}}{\sqrt{f S_{\mathrm{n}}(f)}}  \ ,
\label{eq:snroriginal}
\end{equation}
in which $h_{A}$ is the (time-domain) strain amplitude of the GW signal, $\sqrt{fT}$ is the number of observed cycles, and $S_{\mathrm{n}}(f)$ is the
spectral noise density of LISA evaluated at GW frequency $f$
\citep[we adopt the LISA-N2A5 noise model, see, e.g.,][]{2016PhRvD..93b4003K,Robson+19}\footnote{In this work, we do not take into account the potential stochastic GW background created by astrophysical bursting sources when evaluating the LISA noise curve, but they may have a significant contribution to the noise level \citep[see, e.g.,][]{Naoz+23}.}, $\sqrt{f S_n(f)}$ is the dimensionless spectral noise amplitude per logarithmic frequency bin. The numerator in Eq.~\eqref{eq:snroriginal} defines $h_c$, the characteristic strain for a monochromatic source.

Note that $T$ in Equation~(\ref{eq:snroriginal}) stands for the time of observation when the signal is present, thus is different from the total observational time of LISA, $T_{\rm obs}$. For bursting sources, we sum over the pericenter passage time when they are emitting GW with significant amplitude (comparable with the corresponding circular source enclosed by the pericenter distance) to get $T$. For example, in a four-year LISA mission, if a bursting source bursts $n$ times, then we have $T=nT_{\rm burst}$, while $T_{\rm obs}=nT_{\rm orb}=4$~yr.

Plugging in Equation~(\ref{eq:time}), (\ref{eq:frequency}), and (\ref{eq:amplitude}) into Equation~(\ref{eq:snroriginal}), then take into account the small amount of time when the source is bursting ($T=nT_{\rm burst}=T_{\rm obs}/T_{\rm orb}\times T_{\rm burst}$ if $T_{\rm obs}/T_{\rm orb}=n\geq 1$, otherwise $T=T_{\rm burst}$ if $T_{\rm obs
}\leq T_{\rm orb}$), we can get an estimate of the bursting source's detectability:
\begin{align}
&{\rm SNR_{\rm burst}} 
\sim
\frac{h_{\rm burst}\sqrt{f_{\rm burst} T}}{\sqrt{f_{\rm burst} S_n\left(f_{\rm burst }\right)}}  \nonumber  \\ 
&\quad\sim
\left\{
\begin{array}{cc}
     \dfrac{h_{\rm burst}\sqrt{f_{\rm burst} T_{\rm o b s}(1-e)^{3 / 2}}}{\sqrt{f_{\rm burst} S_n(f_{\rm burst })}} & \text{if~} T_{\rm obs} \geq T_{\rm orb}\\[3ex]
     \dfrac{h_{\rm burst}}{\sqrt{f_{\rm burst} S_n(f_{\rm burst })}} & \text{if~}  T_{\rm obs} \leq T_{\rm orb}
\end{array}
\right.
\label{eq:SNR}
\end{align}

Note that for the first case ($T_{\rm obs} \geq T_{\rm orb}$) in Equation~(\ref{eq:SNR}), the SNR can also be expressed as:
\begin{equation}
{\rm SNR_{\rm burst}} 
\sim \frac{h_{\rm burst}}{\sqrt{S_n\left(f_{\rm burst }\right)}} \sqrt{T_{\rm o b s}\left(1-e\right)^{3 / 2}}\text{if~} T_{\rm obs}\geq T_{\rm orb} , 
\label{eq:snrnew}
\end{equation}
which shows the dependency of (repeated) bursting sources' SNR on the eccentricity.

Equation~(\ref{eq:snrnew}) estimates the RB sources' SNR. Furthermore, it can be verified against the numerical results of eccentric binaries' detectability \citep[e.g.,][]{Kocsis_2012,Hoang+19}. For example, the SNR of a bursting BBH system with $m_1=m_2=10$~M$_{\odot}$, $a=1$~$\rm au$, $e=0.999$ is estimated to be $\sim 200$ in the Milky Way center, and $\sim 2$ at the distance of the Andromeda. Therefore, Equation~(\ref{eq:snrnew}) implies that stellar-mass mHz bursting sources can be detected in the Milky Way, and are marginally detectable in nearby galaxies.
\begin{figure}[htbp]
    \centering
    \includegraphics[width=3.5in]{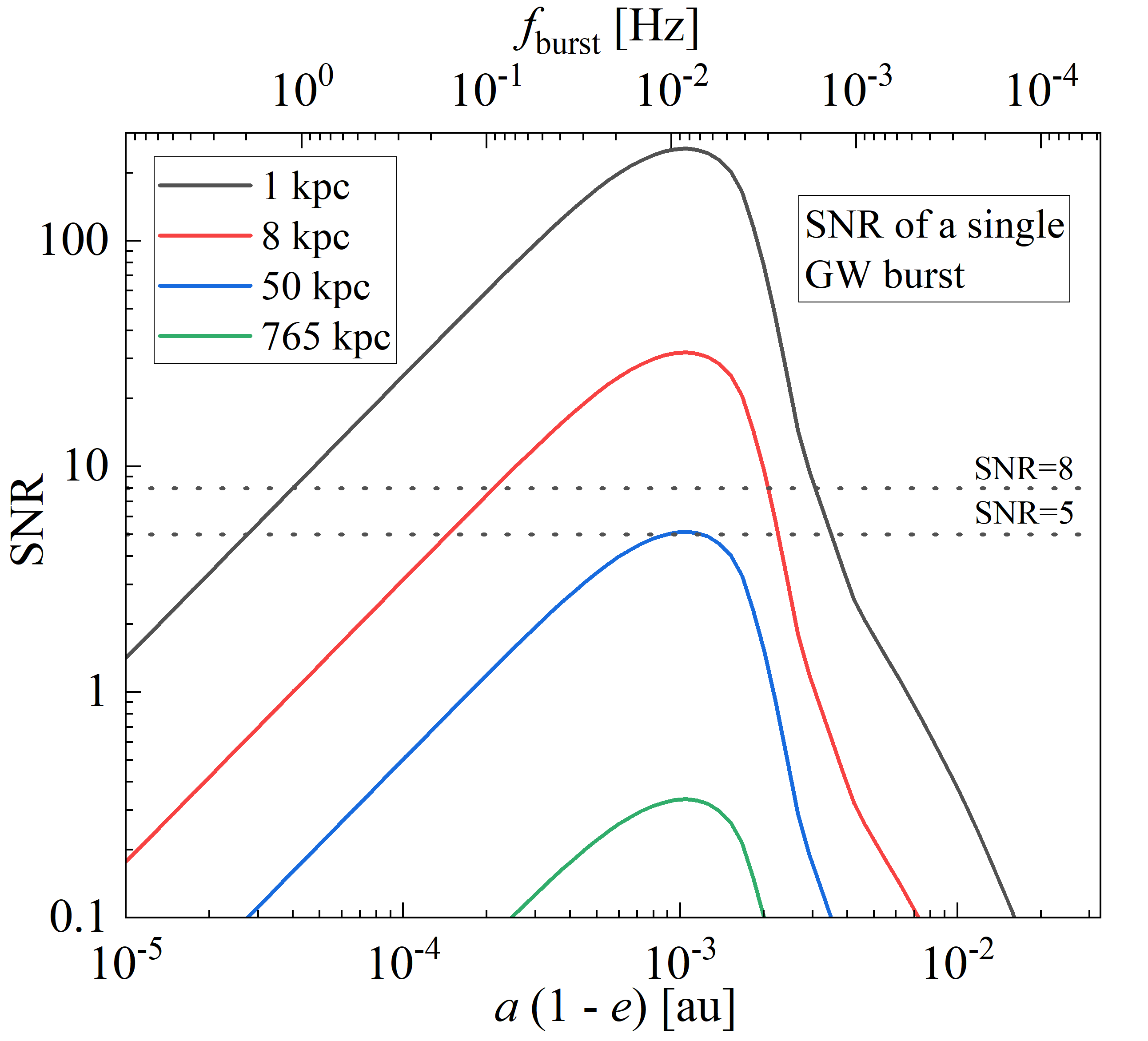}
    \caption{{\bf{The signal-to-noise ratio of a single GW burst, as a function of the source's pericenter distance.}}
    We show a BBH system with 
    $m_{1}= m_{2}=10$~M$_{\odot}$, placed at $D_l=1,8,50,765$~kpc, respectively. The $\rm SNR$ of a single GW burst is plotted as a function of the binary's pericenter distance $a(1-e)$, and the corresponding GW burst frequency is plotted on the top x-axis. We note that repeated burst sources can have multiple bursts detected during the observation, and the overall $\rm SNR$ can be higher than the single burst case of this figure. See Figure \ref{fig:maps} for the repeated burst case.
    }
    \label{fig:singleburst}
\end{figure}

For the second case ($T_{\rm obs} \leq T_{\rm orb}$) in Equation~(\ref{eq:SNR}), the binary can only undergo one pericenter passage during the observation. Therefore, the signal-to-noise ratio represents a single GW burst's detectability. We show in Figure~\ref{fig:singleburst} that such $\rm SNR$ of a single burst can be expressed as a function of the binary's pericenter distance, and for stellar-mass binaries, the distance of single burst detection is mostly limited to the Milky Way. The SNR peaks at a burst frequency of about 6 mHz, the minimum of the LISA sensitivity curve, corresponding to a pericenter distance of $\sim 10^{-3}$~au.

\begin{figure}[htbp]
    \centering
    \includegraphics[width=3.5in]{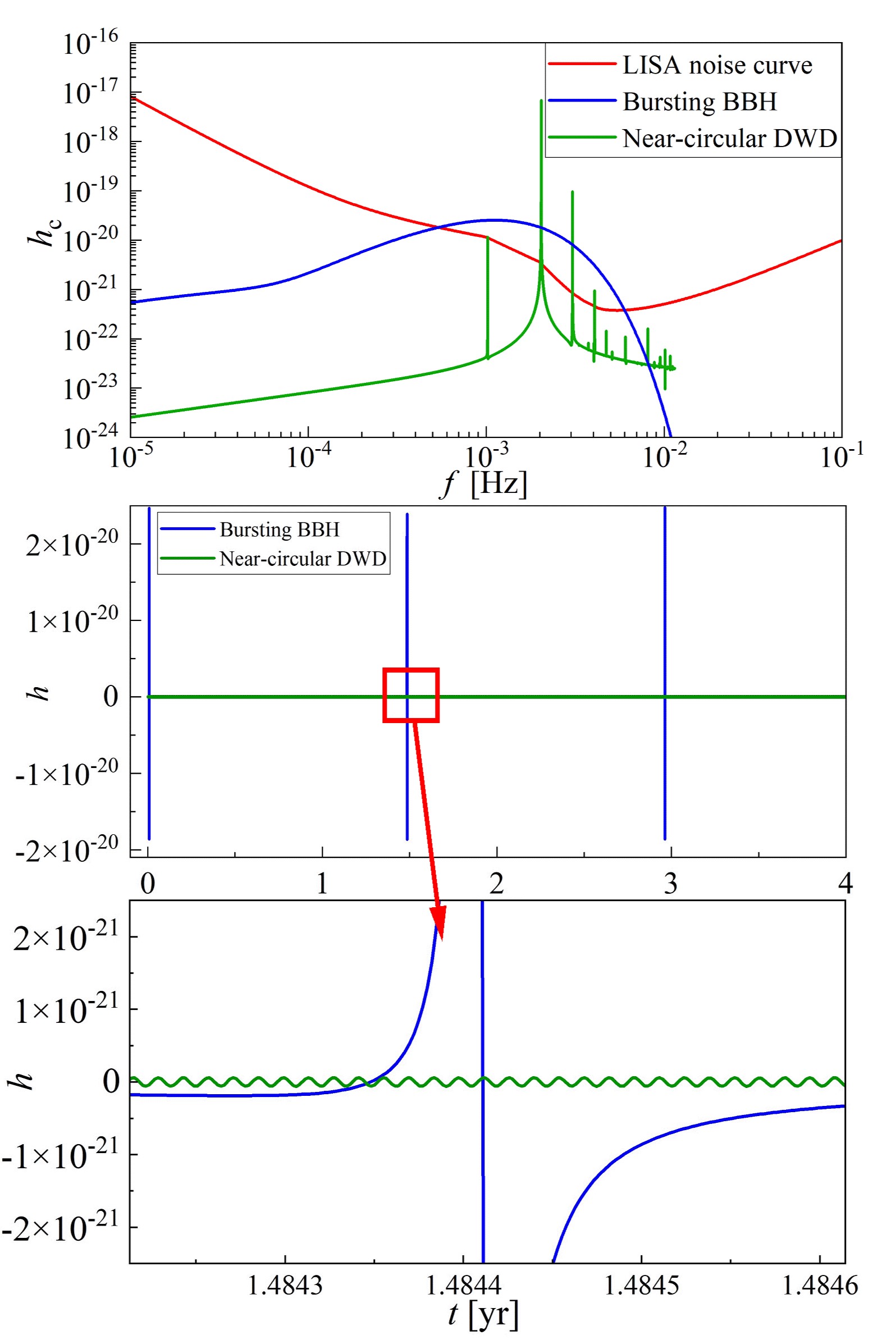}
    \caption{{\bf{Comparison between the GW signal of a bursting BBH and a near-circular DWD, both with the same signal-to-noise ratio $\rm SNR\sim 12$.}}
    We show a BBH system (blue line) with 
    $m_{1}= m_{2}=20$~M$_{\odot},$ 
    $a=4.43$~au,
    $e=0.999$
    and a DWD system (green line) with     $m_{1}= m_{2}=0.5$~M$_{\odot},$ 
    $a=0.001$~au,
    $e=0.01$, both of them are placed at $D_{l}=8$~kpc, and observed for 
    $T_{\rm obs}=4$~yr. 
    These parameters are chosen such that these sources will have
     $\rm SNR=12$. 
    {\it Upper Panel} shows the characteristic strain (Eq.~\ref{eq:hc}) compared with the LISA noise curve ($\sqrt{fS_n(f)}$) (red line), {\it  Middle Panel} and {\it Bottom Panel} show the time-domain waveforms. This plot highlights that bursting signals in the milli-hertz band can have significant amplitude in the time domain (thousands of times greater than a circular source with the same SNR), even if their averaged SNR is limited. 
    }
    \label{fig:LISA2systems}
\end{figure}

We emphasize that, the expression of SNR in Equation~(\ref{eq:SNR}) are based on the average GW emission power, and thus may not describe the full detectability of bursting sources. For example, Figure~\ref{fig:LISA2systems} shows the comparison of time and frequency domain waveforms of a bursting BBH and a circular double white dwarf (DWD), calculated numerically using the x-model \citep{Hinder+10}.\footnote{We note that the Fourier transform of highly eccentric sources' GW signal can be made up of millions of harmonics, in the {\it{Upper Panel}} of Figure~\ref{fig:LISA2systems} we only plot their spectrum density for simplicity. A detailed discussion can be found in \citet{Kocsis_2012}.} For a comprehensive comparison, we adjust their parameters to have 
the same $\rm SNR\sim 12$.
As shown in the figure, even if the bursting BBH system has the same SNR as the circular DWDs, its time-domain GW amplitude is thousands of times greater than the latter one. In other words, with the same SNR, the time-domain strain amplitude of a mHz ``burst" is much greater than that of a continuous GW. Therefore, we may use this signature to enhance the bursting sources' detectability \citep[e.g.,][]{east13,Tai_2014,Loutrel+20,wu2023searching}.

We note, however, that the transient nature of bursting sources also adds to the difficulty of data analysis. In particular, the typical strategy for detecting GWs uses matched filtering \citep{Finn92,Cutler+94}, which heavily relies on the accurate model of GW signal and is not well-suited for searching for discrete bursts localized in time and frequency \citep[see, e.g., ][]{Tai_2014,Loutrel_2017,Loutrel+20}. In the context of LIGO data analysis, extensive efforts were made to identify transient events using time-frequency methods, such as power stacking \citep{east13}, the TFCLUSTER algorithm \citep{sylvestre02}, wavelet decomposition \citep{Klimenko_2004}, and the Q-transform \citep{bassetti2005development,Tai_2014}. These methods may lead to a lower $\rm SNR$ relative to the matched filtering but are robust to different kinds of transients and can combine the information from multiple bursts from a single source. For LISA data analysis, burst detection methods for stellar mass binaries are still underdeveloped, but some efforts have been made to detect highly eccentric EMRIs  \citep[see, e.g., ][]{barack04,Cornish+03,Hopman_2007,porter2010eccentric,mikoczi2012} and burst from scattering of black holes \citep{Kocsis_2012}. 

Moreover, the detection of bursts with the LIGO-VIRGO-KAGRA network has the advantage that the source may be more precisely localized in the sky, given the ability to measure the GW arrival time difference at different detector sites. In contrast, for persistent inspiraling sources in the LISA band, their localization depends on the modulation of the signal caused by the annual motion of the LISA satellite constellation along its orbit. While ``time delay interferometry" \citep[see, e.g.,][]{Tinto+04,Tinto_2021} may be utilized to identify unmodeled astrophysical GW sources with LISA, corresponding mock data analysis methods are currently unavailable for cases with only a small number of bursts in the observational period. Thus, it is currently not known how accurately the source may be localized in the sky, if at all in this case. Additionally, distance information may also be unavailable, as it degenerates with the total mass of the source and the separation at the closest approach. Thus, confusion of several distinct binaries with repeating single GW bursts may be a major uncertainty for these types of sources. It remains unclear how many repeated bursting sources (with what SNR) are needed for secure detection. For conservative purposes, in this paper, we adopt the average SNR as the criteria for evaluating the bursting sources' detectability. This approach represents the power of the GW signal and provides an optimistic estimate of bursting sources' detectability in the LISA band with current data analysis methods.

\subsection{Lifetime of Bursting Sources in the Milli-Hertz Band}
\label{subsec:lifetime}
Most of the stellar-mass bursting sources start their evolution from a wide configuration. These systems undergo some dynamical interaction, such as EKL, fly-by, or a strong encounter, resulting in a highly eccentric configuration. At this stage, the system undergoes repeated GW bursts. Following this RB stage, the binary's orbit shrinks, and if GW emission dominates the evolution, it becomes circularized, yielding a merged system (e.g., see Figure~\ref{fig:EKLevolution}). 
Particularly, an RB source's lifetime, $\tau_{\rm RB}$, can be estimated using the merger timescale for binaries with extreme eccentricity \citep{Peters64}:
 
\begin{align}
    \tau_{\mathrm{RB}}\sim & \frac{3}{85\mu M^2} a^4\left(1-e^2\right)^{7 / 2} 
    \sim  1.12\times10^{7}{\rm yr}\times \nonumber\\&\,\eta_s^{-1}\left(\frac{m}{10\rm M_{\odot}}\right)^{-\frac{5}{3}} \left(\frac{f_{\rm burst}}{1\rm mHz}\right)^{-\frac{8}{3}}\left(\frac{1-e}{0.01}\right)^{-\frac{1}{2}} ,
    \label{eq:lifetime}
\end{align}
where $\mu=m_{1}m_{2}/(m_1+m_2)$, and $M=m_1+m_2$.

Equation~(\ref{eq:lifetime}) describes the isolated evolution of bursting binaries. However, some bursting sources that undergo dynamical interaction can have eccentricity oscillation throughout their evolution \citep[e.g., EKL merger systems][]{Hoang+19,Randall+19,Deme+20,Emami+20,Chandramouli+21}. In this case, we adopt the maximum eccentricity during the oscillation, $e_{\rm max}$, as the value of $e$ in Equation~(\ref{eq:lifetime}), which yields a lower limit of the timescale for a source to burst.

We are interested in comparing the length of a bursting system's RB stage with its latter stage of evolution (inspiral with moderate eccentricity). Thus, for simplicity, we assume that the binary circularizes and shrinks to a radius $r_c\sim a(1-e)$ after the RB evolution. Such systems are well described in EKL systems or captured systems \citep[e.g., see][]{Kocsis_2012,Naoz16}. Further, this approximation is consistent with the concept that the pericenter of a highly eccentric GW source remains nearly the same \citep[in the absence of dynamical evolution, see for details ][]{peters63,Peters64}.

Thus, under this notion, when the bursting binary's eccentricity drops to a moderate value, its remaining inspiral time, $\tau_{\rm inspiral}$, can be estimated as \citep{Peters64}:
\begin{align}
\tau_{\rm inspiral} &\sim \frac{5}{256\mu M^2}r_c^4 \sim \frac{5}{256\mu M^2}a^4(1-e)^4 \nonumber\\
&\sim  5.5\times10^{4} {\rm yr} \,\eta_s^{-1} \left(\frac{m}{10 \,\rm M_{\odot}}\right)^{-\frac{5}{3}}\left(\frac{f_{\rm burst}}{1\,\rm mHz}\right)^{-\frac{8}{3}}
\ .
\label{eq:inspiral}
\end{align}

We note that Equation~(\ref{eq:inspiral}) also serves as a general estimate of millihertz circular BBHs' merger timescale (if we replace $f_{\rm burst}$ in the equation with $f_{\rm circ, GW}$). Therefore, the equation implies that a stellar mass circular binary will stay in the $\rm mHz$ band, in particular $f_{\rm circ, GW}\gtrsim 1\rm mHz$, for $\sim 10^3-10^5$~yr \citep[e.g., ][]{chen20gas}.

As can be seen from Equation~(\ref{eq:lifetime}) and (\ref{eq:inspiral}), the timescale for a highly-eccentric source's RB stage is longer than the following inspiral stage with moderate eccentricity:
\begin{equation}\label{eq:RBtime}
    \tau_{\rm RB}\sim 20  (1-e)^{-\frac{1}{2}} \tau_{\rm inspiral}\ ,
\end{equation}
which reflects the fact that in the RB stage, the eccentric binary will spend most of its time at large separation, yielding a lower 
GW emission and orbital energy loss compared to a circular orbit with a separation $a(1-e)$. For example, the EKL mergers of BBHs in the galactic nucleus are expected to have  $e\gtrsim 0.999$, induced by the supermassive black (SMBH) via the EK mechanism \citep{Naoz16,Hoang+18,Hoang_2019}. Thus, Equation (\ref{eq:RBtime}) means the RB stage of EKL mergers will be hundreds of times longer than their inspiral time with moderate eccentricity and the same $a(1-e)$.
\begin{figure*}[htbp]
    \centering
    \includegraphics[width=3.5in]{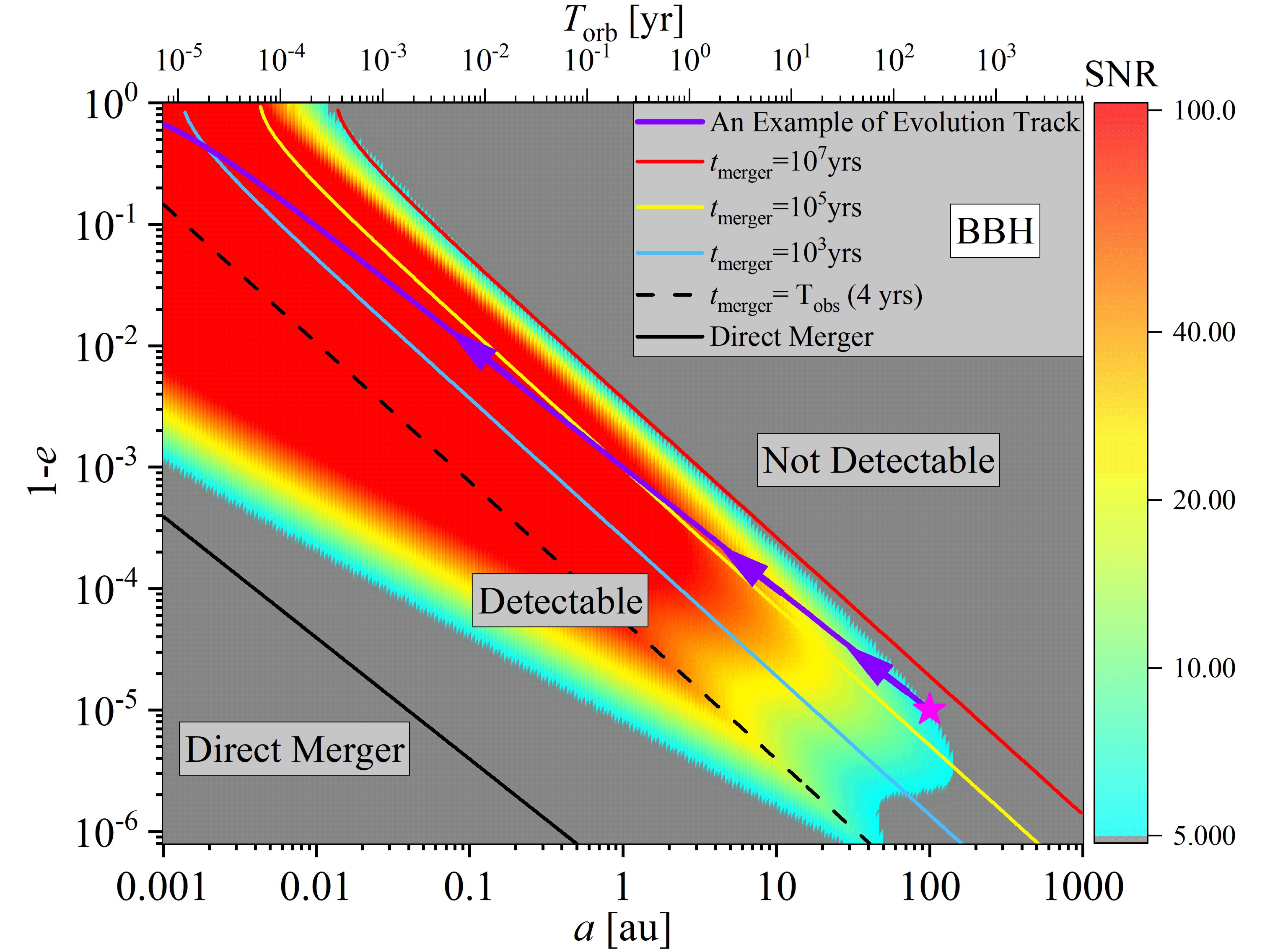} \includegraphics[width=3.5in]{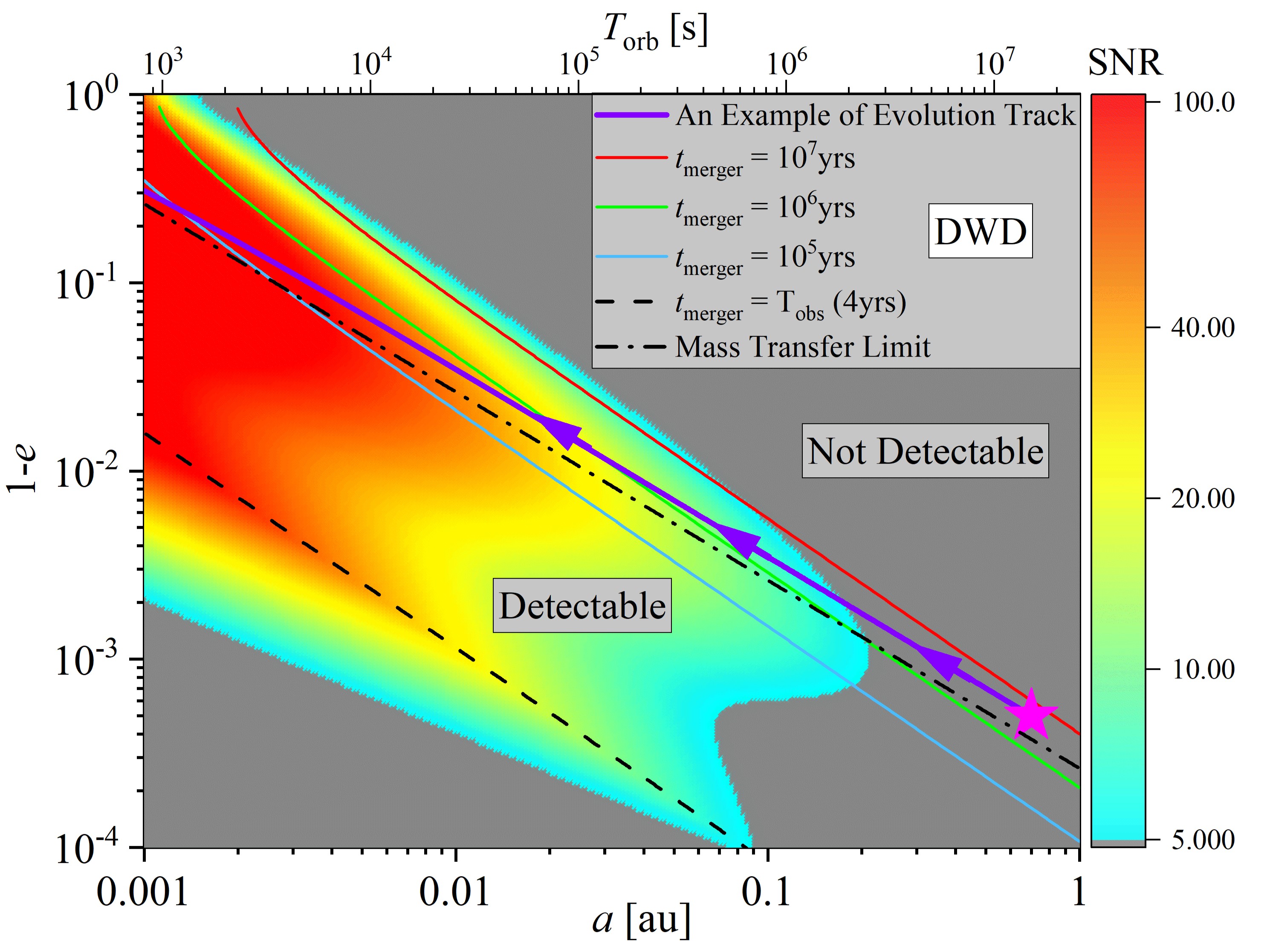}
    \caption{{\bf{The Dependence of the (RB) compact binary's detectability on its semi-major axis and eccentricity. }}
    Here we show a BBH system ({\it Left Panel}) with $m_{1}= m_{2}=10$~M$_{\odot},$ and a DWD system ({\it Right Panel}) with $m_{1}= m_{2}=0.75$~M$_{\odot},$ placed at
    $D_{l}=8$~kpc and observed for $4$~yr with LISA. The color represents the compact binary's signal-to-noise ratio as a function of its semi-major axis $a$ and eccentricity $1-e$. 
    As a proof of concept, we depict 
    an example system's evolution (purple line with star and arrows) for both parameter spaces and plot the corresponding orbital time on the top x-axis. 
    The solid lines with color represent the merger timescale $\tau_{\rm merger}$ for a given $(a,e)$ configuration of the binary. In other words, when the evolution track of the binary gets into the colored region and gets across a given merger timescale, we can estimate the remaining time for it to emit detectable GW bursts before the merger. We note that the signal-to-noise ratio is suppressed in the region below the dashed lines ($t_{\rm merger} < T_{\rm obs}$) because the binary will merge during the observation, thus having less observational time than the total LISA mission time. Therefore, when calculating the SNR for these binaries following Equation.~(\ref{eq:snrsum2}), we replaced the $T_{\rm obs}$ with $t
_{\rm merger}$.
    }
    \label{fig:maps}
\end{figure*}

 Figure~\ref{fig:maps}
 shows the lifetime (Eq.~\ref{eq:RBtime}) and detectability of RB sources with a more detailed calculation for the SNR than in Equation (\ref{eq:SNR}), as depicted in Appendix~\ref{append:snr}. Particularly, we consider equal mass $10$~M$_{\odot}$ BBH systems (left panel), and equal mass $0.75$~M$_{\odot}$ DWD systems (right panel) at a luminosity distance $D_{l}=8$~kpc from the detector. The Figure shows the GW signal-to-noise ratio, as a color map of the binary's semi-major axis $a$ and  $e$. The detectable regions are labeled on the map.

We also over-plot, in Figure~\ref{fig:maps}, the merger timescales for isolated binaries in each panel (thin solid lines in the figure). For comparison purposes, in the left panel, we further show the direct merger limit for BBHs (black solid line at the bottom). 
In the right panel, we show the Roche limit for DWD in the black dash-dotted line (which is the limit that DWDs will undergo mass transfer). 

Overplotted in Figure \ref{fig:maps} is a representative example. Specifically, starting from a given initial condition, the RB source will evolve in the $a-e$ parameter space following the changes in its semi-major axis and eccentricity due to the GW emission \citep{Peters64}:
\begin{equation}
\left\langle\frac{d a}{d e}\right\rangle=\frac{12}{19}
\frac{a}{e}\frac{\left[1+(73 / 24) e^2+(37 / 96) e^4\right]}{\left(1-e^2\right)\left[1+(121 / 304) e^2\right]} \ .
\label{eq:isolated ae}
\end{equation}
Using these equations, we plot the evolutionary track (the solid lines with arrows) on each panel in the absence of further dynamical evolution. 

Consider, for example, the evolutionary track in the right panel. This DWD system began in the undetectable regime; however, as it evolves, it becomes detectable via RB, with SNRs corresponding to the color on the map. When the system crosses the lifetime line, it will merge within that time, thus undergoing repeated bursts for that remaining time. In particular, the DWD example crosses the $10^6$~yr mark with $a\sim 0.03$~au and $e\sim 0.99$, which means it will survive for an additional $10^{6}$ years as a detectable repeated burst source. 

As can be seen in Figure~\ref{fig:maps}, the lifetime of a detectable RB source in the Milky Way can be as much as $\sim 10^{7}$~yr, since the edge of the detectable region roughly coincides with the line of merger timescale $t_{\rm merger}=10^{7}$~yr for both BBHs and DWDs. 
This phenomenon can be explained using Equation~(\ref{eq:snrnew}) and (\ref{eq:lifetime}). Particularly, for extreme eccentricity cases, the merger timescale lines follow $a^4(1-e)^{7/2}=\rm constant$ in the a-e parameters space (see Equation~(\ref{eq:lifetime})), while the edge of the detectable region (SNR=5 line) follows $h_{\rm burst}(1-e)^{3/4}(S_n(f_{\rm burst}))^{-1/2}=\rm constant$ (see Equation~(\ref{eq:snrnew})). For the expression of the constant SNR contour, we can substitute $h_{\rm burst}\sim a^{-1}(1-e)^{-1}$ (see Equation~(\ref{eq:amplitude})) and $f_{\rm burst}\sim a^{-3/2}(1-e)^{-3/2}$ (see Equation~(\ref{eq:frequency})), then take into account the frequency dependence of the LISA noise curve (e.g., for N2A5 configuration, dominant term $S_n(f)\sim 20/3f^{-4.4}\times10^{-50.92}$ in the millihertz band \citep{Klein+16}) and get that $a^{-4.3}(1-e)^{-3.55}\sim \rm constant$. Therefore, in Figure~\ref{fig:maps}, the constant $t_{\rm merger}$ lines (${\rm log}_{10}a+0.88{\rm log}_{10}(1-e)\sim \rm constant$) nearly follow the same $a-e$ dependence as the constant SNR lines (${\rm log}_{10}a+0.83{\rm log}_{10}(1-e)\sim \rm constant$), indicating that each SNR level roughly corresponds to a given merger timescale. In our example, the specific values of SNR=5 and $t_{\rm merger}=10^{7}$~yr are coincidental. If certain dynamical formation channel creates compact binaries that evolve and get across this region, they will become long-lived mHz GW sources ($\tau_{\rm RB} \sim 10^{6}-10^{7}$~yr, see Equation~(\ref{eq:lifetime})) with a detectable time much longer than the circular inspirals within the same frequency band ($\tau_{\rm inspiral} \sim 10^{3}-10^{5}$~yr, see Equation~(\ref{eq:inspiral})). 

We note that in Figure~\ref{fig:maps}, another population of BBHs potentially exists, which is characterized by moderate eccentricity ($e\sim0$) and long lifetime ($t_{\rm merger}\sim 10^7$~yr). However, these systems exhibit a much lower GW frequency ($\sim 0.1\rm mHz$) compared to the bursting sources considered in this paper ($\gtrsim 1\rm mHz$). For example, a BBH system with $e\sim0, a\sim 0.03$~au would be detectable in the Milky Way ($\sim 10\rm kpc$) for $10^7$~yr, with an orbital period of $T_{\rm orb}\sim 10^{4}$~s and GW frequency $f_{\rm circ,GW}\sim 10^{-4}$~Hz. Note that because we expect a large population of circular DWDs in the sub-millihertz band \citep[see, e.g.,][]{Nissanke12, lamberts18,Xuan+21}, the identification of the near-circular, low-frequency BBHs is beyond the scope of this study. Thus, in this paper, we focus on the dynamically formed, highly-eccentric bursting sources, with $f_{\rm GW}\gtrsim 1\rm mHz$.

\begin{figure*}[htbp]
    \centering
    \includegraphics[width=3.5in]{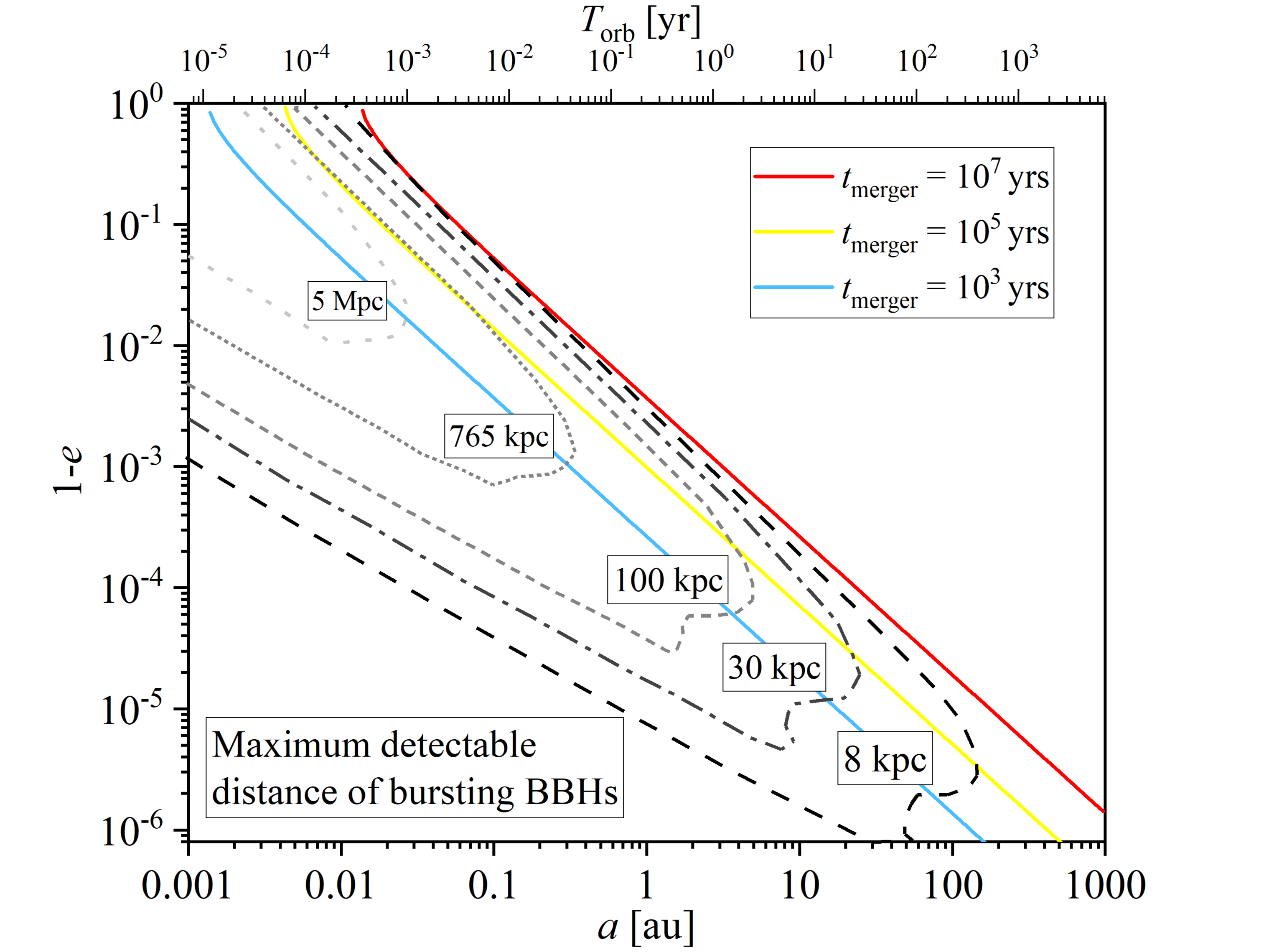} \includegraphics[width=3.5in]{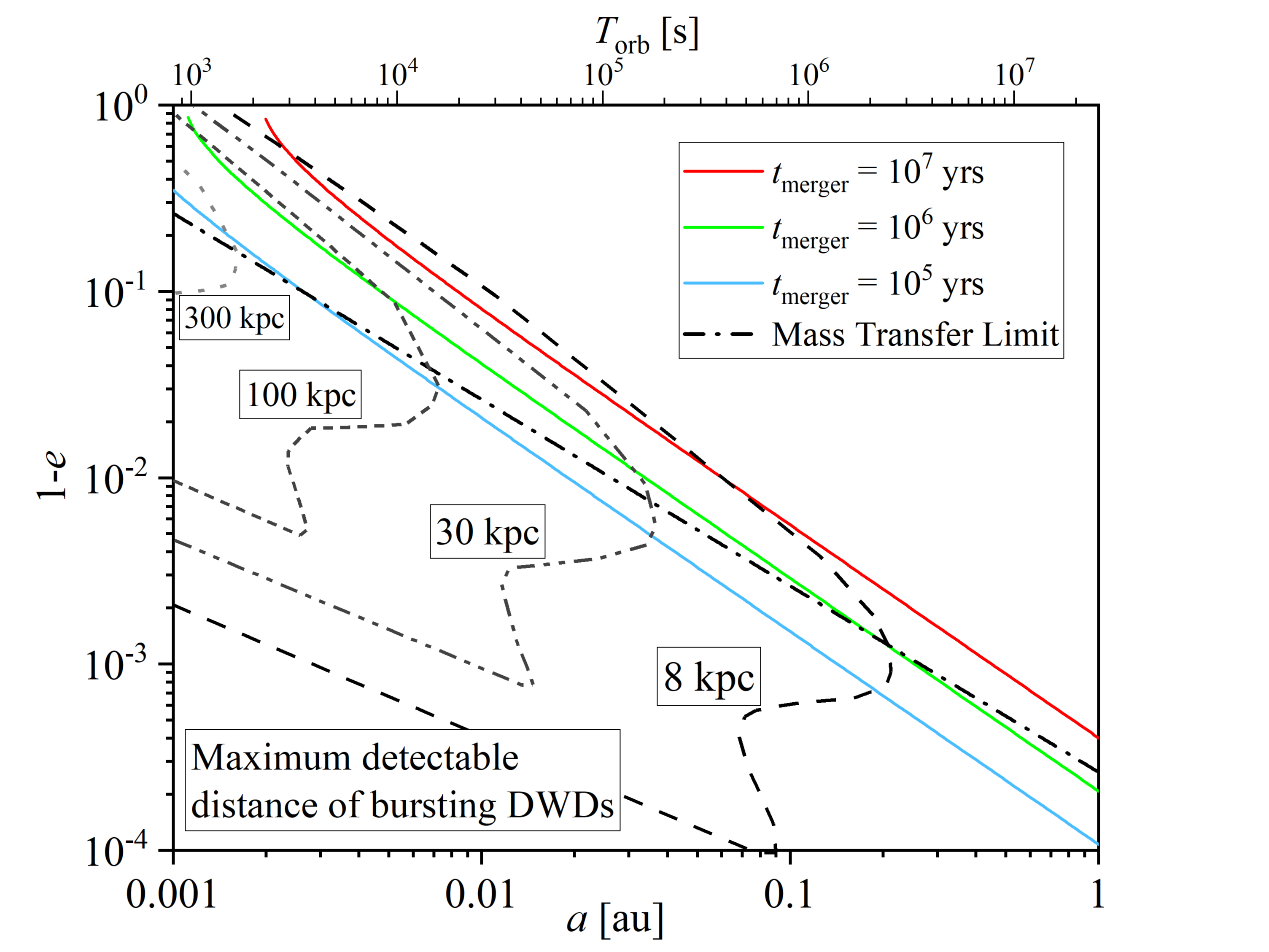}
    \caption{{\bf{The maximum detectable distance of the (RB) compact binary. }}
    Here we consider the same systems as in Figure~\ref{fig:maps}, but choose different values of luminosity distance $D_{l}$, and show the equal signal-to-noise ratio ($\rm SNR=5$) contours for a $4$~yr LISA observation. The left (right) panel shows the BBHs (DWDs) system. The detectability of bursting sources increases at the top-left part of the plot. 
    }
    
    \label{fig:maps2}
\end{figure*}

For comparison purposes, we show in Figure~\ref{fig:maps2} the maximum detectable distance of RB sources, assuming a $4$~yr LISA mission and a detection threshold of $\rm SNR=5$. As shown in the figure, the detection of bursting BBHs can be promising within the distance of nearby galaxies, while the detection of bursting DWDs is limited to the Milky Way. This result is consistent with the estimation in Section~\ref{subsec:detectability} (see, Equations~(\ref{eq:SNR}), (\ref{eq:snrnew}) and Figure~\ref{fig:singleburst}).

One important caveat is that the estimation of SNR in Figure~\ref{fig:maps} and \ref{fig:maps2} is based on the averaged power of the GW signal, where
for a binary with $T_{\rm orb}$ larger than $T_{\rm obs}$, the number of bursts we expect to detect during the observation is smaller than one. In this case, the average SNR in Figure~\ref{fig:maps} is smaller than the SNR of detected bursts because we detect only a fraction of such bursting sources, only having a probability of less than one for detection. However, for these non-repeated bursting sources, a single burst falling within the observational window will be brighter than shown in the Figure according to Eq.~\eqref{eq:snrnew}, and hence can still be bright enough to detect with LISA. Moreover, in the region below the dashed black line of Figure~\ref{fig:maps}, the merger timescale of a GW source is shorter than the observational duration, which means the fast-merging sources in this region will not be observed for the whole LISA mission, thus having lower SNR. Therefore, we can further include these facts and finally get:
\begin{equation}
{\rm SNR} \sim \frac{h_{\rm burst}\sqrt{\max\{\min(\tau_{\rm insp},T_{\rm obs}),T_{\rm orb}\}(1-e)^{3 / 2}}}{\sqrt{S_n\left(f_{\rm burst }\right)}} .
\label{eq:SNR2}
\end{equation}
in which $\tau_{\rm insp}$ is the binary's remaining inspiral time.

Equation~(\ref{eq:SNR2}) gives the full detectability of (both repeated and non-repeated) burst sources, providing that at least one burst takes place during the observation. The suppression of SNR caused by $\tau_{\rm insp}$ is taken into account in the numerical results of this paper.


\section{Population Estimation and Astrophysical Consequences}
\label{sec:population}

\subsection{General Considerations}
As shown in Section.~\ref{sec:RB property}, the RB stage of a stellar-mass GW source is well detectable within $\lsim 10$~kpc, and lasts longer than the inspiral stage with moderate eccentricity. Therefore, the population of bursting sources can have a significant contribution to the detection of dynamically formed GW sources, especially in the Milky Way.

\begin{figure}
    \centering
    \includegraphics[width=3.5in]{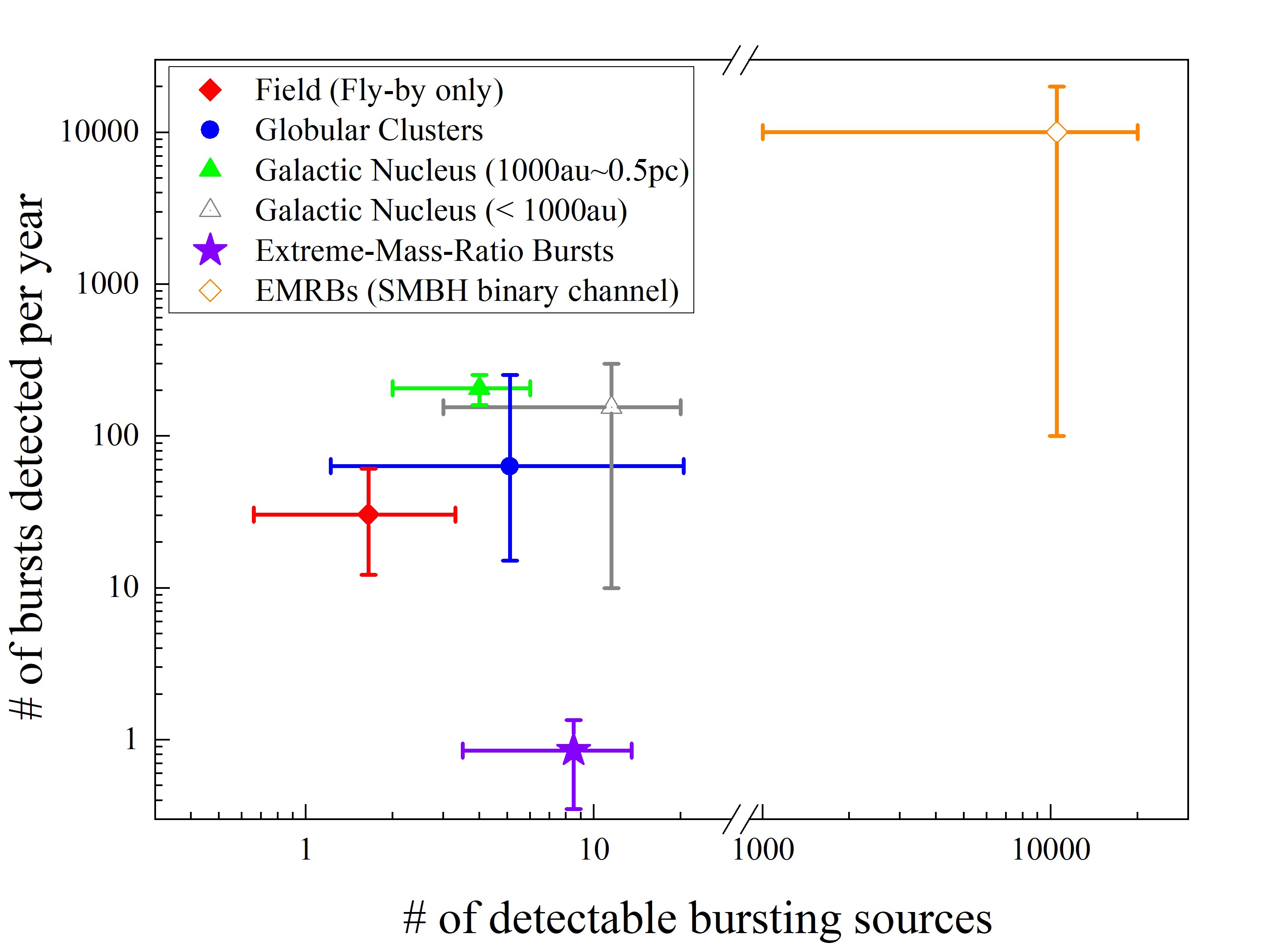}
    \caption{{\bf The number of bursting sources and bursts detected per year, assuming a 10-yr LISA mission and detection threshold $\rm SNR = 5$. } The figure depicts the estimation results of bursting BBHs for different channels, with the error bars reflecting the theoretical uncertainties. Specifically, we show the number of detectable bursting sources from each channel (x-axis) and their contribution to the number of bursts detected per year (y-axis).  
    For details of the simulation, see Appendix~\ref{append:details}. 
    Note that the number of detectable bursting sources is roughly proportional to the observational time. The solid points represent the population with convincing evidence of existence, while the hollow points are for the potentially existing population of bursting sources. We emphasize that if future detection does not show evidence of the population marked in hollow points, it serves as a strong constraint on the corresponding formation channels.
    }
    \label{fig:population}
\end{figure}

As a proof of concept, we heuristically estimate the expected populations of bursting BBHs. Due to the large uncertainties of star formation and detailed dynamical evolution, we focus on the expected order of magnitude of the bursts. We consider three regimes that are expected to have eccentric sources, namely: Galactic Field, Globular Clusters (GCs), and Galactic Nucleus (GN), and show the results in Figure~\ref{fig:population}. For comparison purposes, we also present the number estimation of GW bursts from the EMRIs in the universe.

For other kinds of bursting sources, such as DWDs, their number, parameter distribution, and evolution is less understood, mostly due to the uncertainty in the formation scenario and the complexity of tidal effects in extreme eccentricity cases. Therefore, we will leave these populations for future work.

Before providing the estimation for bursting sources in the three regions mentioned above, we first describe a general estimate agnostic to the astrophysical formation channel of the source. 

\subsection{Steady-State Approximation and Constraints from LVK}
\label{sec:ligoconstrain}
Motivated by LVK observations, we first estimate the number of RB sources under the assumption that LVK represents a steady-state population. In particular, assume there is a continuous birth and death of compact object binaries in the Milky Way, keeping the total number of systems unchanged. In this case, the replenishing rate of a GW source is balanced by its merger rate. In other words, 
\begin{equation}
    \Gamma_{\rm rep}=\Gamma_{\rm merger} \ .
\end{equation}
For GW sources formed through dynamical interaction, the expected number of systems in the RB stage, $N_{\rm RB}$, is found by multiplying the average lifetime of the RB stage (Eq.~(\ref{eq:lifetime})) with the formation rate of the source:
\begin{equation}
    N_{\rm RB}=\tau_{\rm RB}\Gamma_{\rm rep, RB}=\tau_{\rm RB}\Gamma_{\rm merger,RB} \ ,
    \label{eq:steadystate}
\end{equation}
which is also applicable to other evolutionary stages, such as the inspiral with moderate eccentricity.

The LVK observations indicate that BBHs merge at a rate of $15–38\, {\rm Gpc^{–3} yr^{-1}}$ \citep{Abbott_2021}. Therefore, we can estimate the BBHs merger rate per galaxy by dividing the galaxy number density. 
Adopting the galaxy number density of $0.02~\rm Mpc^{-3}$ \citep{Conselice_2005}, the merger rate of BBHs in an averaged galaxy is estimated as:
\begin{equation}
    \Gamma_{\rm merger, avg}\sim 10^{-6}\, {\rm yr^{-1}} \ .
    \label{eq:rate}
\end{equation}
Thus, plugging Equation (\ref{eq:rate}) into Equation (\ref{eq:steadystate}) places a constraint on the number of bursting BBHs we can detect locally. For example, bursting GW sources can spend up to a few $10^7$~$\rm yr$ in the detectable regime (Figure~\ref{fig:maps}). Therefore, the upper bound of RB BBHs' number, from the steady-state approximation and LIGO's constraint, is on the order of $N_{\rm RB}\sim  10^{7}\,{\rm yr \, 10^{-6}\, {\rm yr^{-1}}} \sim 10$ sources in the Milky Way. Below we use this approach to place constraints on different formation channels. 

We emphasize that this constraint only serves as a rough estimation of bursting BBHs' number. It can be well-exceeded if taking into account the non-equilibrium formation history of compact binaries. For example, different regions in a galaxy may undergo starbursts while others have a quiescence phase, making the number of bursting sources fluctuate in a wide range. (See Section~\ref{sec:GN} for a detailed analysis.)
Further, assuming that the observed rate of mergers all form at mHz or smaller frequencies\footnote{{Note that if all sources form at higher frequencies during close encounters, e.g. at $0.01\mathrm{Hz}$, it is in principle also possible not to have any mHz sources without violating the LIGO/VIRGO merger rate.}} due to their long lifetime, bursting binaries could be the dominant source in the local mHz BBHs population (see Equation~(\ref{eq:steadystate})). For example, consider the lifetime of circular (non-bursting) BBHs in the mHz band, $\tau_{\rm inspiral} \sim 10^{3}-10^{5}$~yr (see Equation~(\ref{eq:inspiral})). Provided that both bursting and non-bursting sources are detectable within the Milky Way, the shorter lifetime of circular sources results in their small number in the local population, $N_{\rm circ, BBHs} \lesssim 10^{5}\,{\rm yr \, 10^{-6}\, {\rm yr^{-1}}}\sim 0.1$ in the mHz band, while bursting sources represent the majority of mHz sources ($\rm N_{RB}\sim 10$). 

Below, we consider three characteristic BBHs formation channels and provide heuristic estimations for their bursting properties. 

\subsection{BBHs in the Field}
\label{sec:field}

Isolated stellar binaries in the galactic field have been proposed as a possible channel explaining the LVK observations \citep[e.g.,][]{demink15,dominik15,belczynski16,Eldridge17,Giacobbo18,Olejak20}. 
However, in the context of bursting systems, we focus on wide binaries in the field, which are less likely to be seen with LVK without external perturbations. Providing that external perturbations such as flybys or galactic tides will affect their evolution, these wide-field binaries can be driven to extreme eccentricity, emit bursting GW signals, and become mergers in the end. 

We adopt the model proposed in \citet{Michaely+22}, as well as the steady-state approximation, to calculate the eccentricity and semi-major axis distribution for the BBH mergers in the disk of the Milky Way. In particular, following \citet{Michaely+19}, the merger rate from the wide-binaries field channel can be within the range of $5_{-3}^{+5} \mathrm{Gpc}^{-3} \mathrm{yr}^{-1}$ in the local universe (up to $50 \mathrm{Gpc}^{-3} \mathrm{yr}^{-1}$ if taking elliptical galaxies into account, see Michaely et al. in prep). Based on this rate estimate, assuming a optimistic LISA observation timescale $T_{\rm LISA,obs}=10$~yr and detection threshold SNR$=5$, we find that the minimum (maximum) number of detectable bursting BBHs in the MW, induced by fly-by interaction, is $0.7$ ($3.3$), and they are expected to emit $12$ ($61$) GW bursts per year. In these bursting sources, we identified the lower (upper) bound of the repeated burst sources number as $0.4$ ($1.8$), and $0.3$ ($1.5$) for non-repeated burst sources. However, because of the long orbital period of non-repeated burst sources, the chance that we will observe GW bursts during their pericenter passage is small. Thus, the number expectation of non-repeated bursts is negligible ($\sim 0.1$). This is consistent with the estimate in \citet{Kocsis+2006}. See Appendix~\ref{append:field} for more details. 

\subsection{BBHs in Globular Clusters}
\label{sec:gcs}
BBHs formed through dynamical interactions in globular clusters are suggested to be one of the main sources of GW mergers \citep[e.g.,][]{Orazio+18,Samsing+18,Antonini+19,Fragione+19,Martinez+20,Kremer_2020}. Thus, given the $\sim150$ Milky Way's GCs  \citep[e.g.,][]{Harris+96,Baumgardt+18}, we expect a significant number of bursting sources from GCs. Furthermore, it was recently pointed out that BBHs in GCs can have non-negligible eccentricity \citep[even potential detected by LVK][]{Orazio+18, Antonini+19,Zevin_2019,Samsing+19,Martinez+20,Kremer_2020}.

Here, we adopt the eccentricity distribution of BBHs in \citet[][see their figure 4]{Martinez+20} and the spatial distribution of GCs in the Milky Way from \citet{Arakelyan_2018}. Since we are interested in the RB stage where BBHs are in a wide configuration with low GW frequency ($\sim \rm mHz$), we evolve the systems shown in \citet{Martinez+20}, which is in a higher frequency band, back to the former RB stage. The bursting time of these sources is calculated by counting the time difference between the point when they become detectable and the point when their eccentricity drops below $0.9$ (see the example tracks in Figure \ref{fig:maps}). In the simulation, we use steady-state approximation to calculate the expectation of bursting sources number (For detailed information, see Appendix~\ref{append:gcs}).

We adopt the BBH merger rate in GCs to be:
$\mathcal{R}_0=7.2_{-5.5}^{+21.5} \mathrm{Gpc}^{-3} \mathrm{yr}^{-1}$ \citep[e.g.,][]{Rodriguez2016,Kremer_2020,Antonini_2020}. Assuming $T_{\rm LISA,obs}=10$~yr and SNR=$5$, we find that the minimum (maximum) number of detectable bursting sources in the Milky Way GCs is $1.2$ ($20.1$), which corresponds to a number of GW bursts $15.2$ ($253$) per year. Among these sources, the minimum (maximum) number of repeated burst sources is $0.34$ ($5.7$), while the minimum (maximum) number of non-repeated burst sources is $0.85$ ($14.4$). 

\subsection{BBHs in the Galactic Nucleus}
\label{sec:GN}
The Milky Way's galactic nucleus offers a natural place for the formation of bursting BBHs \citep[see, e.g.,][]{Kocsis_2012,Hoang+19,Stephan+19,Arca+23,Zhang24}. In particular, BBHs orbiting around the supermassive black hole in the galactic nucleus will undergo eccentricity excitation via the EKL mechanism, resulting in the bursting signatures on their GW signal. 

The orbital evolution of BBHs in the GN is quite different from the isolated binaries (e.g., Equation~(\ref{eq:isolated ae})) since they are strongly affected by the gravitational perturbation from the SMBH tertiary. Thus, to get the bursting properties of these sources, we carried our detailed simulations of hierarchical triple systems, including the secular equations up to the octupole level of approximation \citep{Naoz+13}, general relativity precession \citep[e.g.,][]{naoz13}, and GW
emission \citep{Peters64,Zwick+20}. These calculations follow 
a similar approach to \citet{Hoang+18}.

We first present the results under the steady-state approximation (see details in Appendix~\ref{append:GN}). This approximation can well-describe the main population of stars in the GN \citep[old population, aged $2\sim 8~\rm Gyr$][]{Chen+23}. Assuming a $10$~yr LISA observation with the signal-to-noise ratio threshold $5$, we got the expectation of EKL induced, bursting BBHs number $\sim 1$ in the mHz band, with the number of detectable bursts $\sim 100$ per year. All the bursting sources in the simulation are repeated burst sources, which is consistent with the fact that wide binaries will evaporate quickly in the active dynamical environment of GN. The result is calibrated using the $m-\sigma$ relation \citep{merritt1999black,Kormendy+13} and observational results of the Milky Way center.

However, unlike the previous channels (i.e., in the field and GCs), we go beyond the steady state approximation because observations suggest a recent ($2\sim 8$ Myr) star formation occurred in the GN \citep{Paumard06,Lu_2008,Lu2009,Do2013,Chen+23}. Specifically, this recent star formation may have formed a young nuclear star cluster (YNC) within $0.5~\rm pc$ from the central SMBH, with a total mass of $\sim(1.4-3.7)\times 10^{4}M_{\odot}$ and a top-heavy mass distribution \citep{Lu2009}. Therefore, it can hold $\sim 100 - 400$ BHs as a result of the stellar evolution. Furthermore, since stars often reside in binaries, and high-mass stars reside in higher multiples \citep{Pribulla+06,Tokovinin+08,Raghavan+10,Sana+11,Sana+12,Moe+17}, we expect these black holes to form $\sim 100$ BBH systems in the YNC. This expectation is supported by myriad observational and theoretical arguments \citep[e.g.,][]{Ott1999,Martins+06,Rafelski+07,Pfuhl+14,Alexander+14,Naoz+18,Gautam+19,Chu+23}. 
\begin{figure}[htbp]
    \centering
    \includegraphics[width=3.5in]{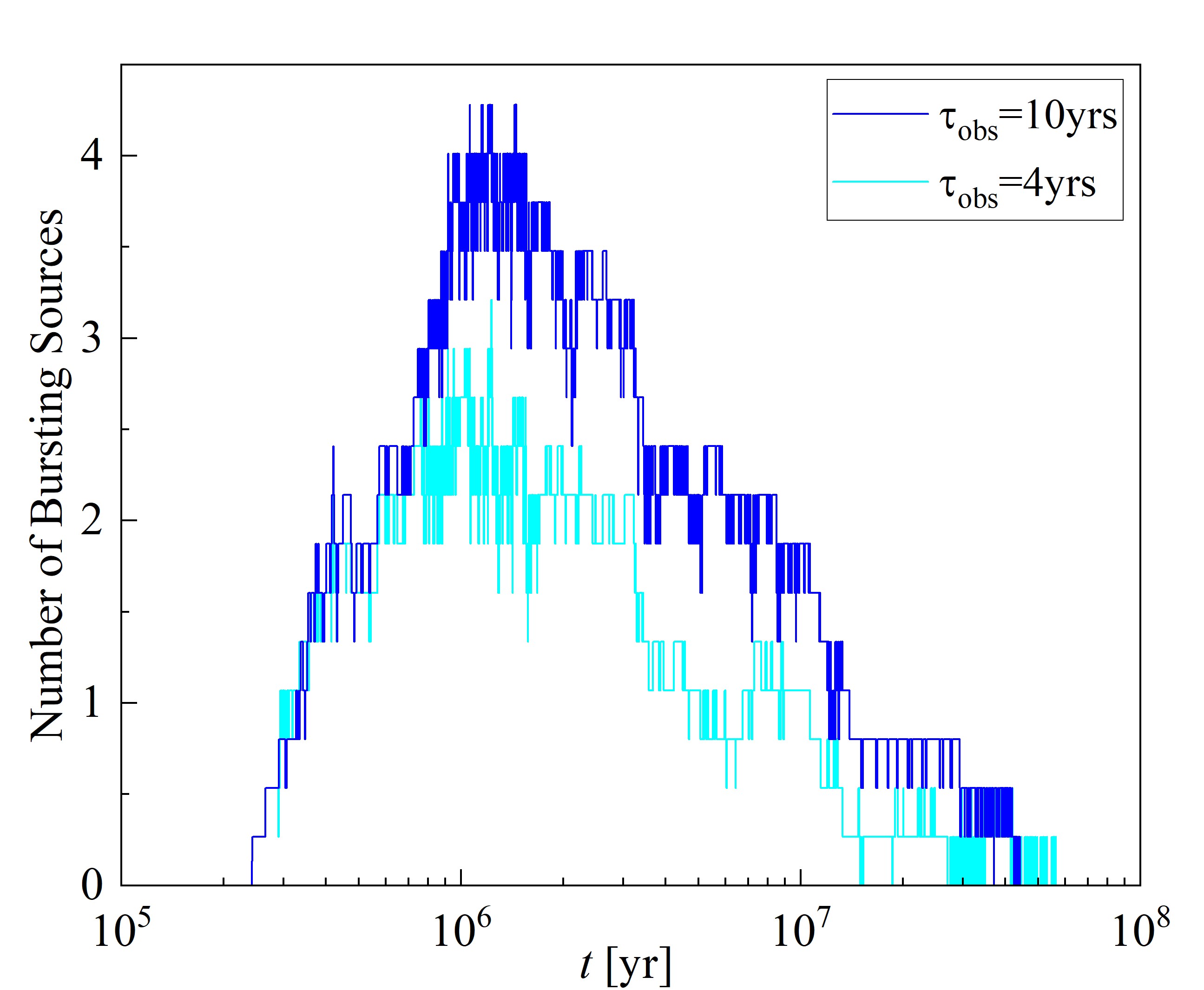}
    \caption{{\bf{Number of detectable bursting sources in the galactic nucleus YNC, as a function of their age.           }}
    Here we show the expectation of observable bursting systems in the Milky Way center young nuclear cluster, as a function of the cluster's age. The deep blue line stands for a LISA observational period $T_{\rm obs}= 10 ~\rm yr$ while the light blue line represents $T_{\rm obs}= 4 ~\rm yr$. The YNC's age is constrained by observation to be $2\sim 8 ~\rm Myr$.
    }
    \label{fig:number of source}
\end{figure}
For the newly-born BBHs, we carried out Monte-Carlo simulations to get the evolution of RB sources' number as a function of the age of YNC (see Figure~\ref{fig:number of source}), as well as the number of bursts in the Appendix (see Figure~\ref{fig:appendGN}). As is shown by the simulation, although the YNC has a relatively small mass ($M\sim 10^4 ~\rm M_{\odot}$), there can be $2\sim 4$ RB sources with $60\sim 150$ bursts detected per year. These numbers are larger than the expectation of bursting systems from the old population of stars under the steady-state approximation, even if the latter one has a much larger total mass of stars ($\sim 10^{7}~\rm M_{\odot}$).

Combining the simulation results of bursting BBHs in the old population of stars with those in the YNC, we get the number estimation of detectable bursting BBHs in the inner $1000~{\rm au}\sim 0.5 ~\rm pc$ of the galactic center as $2\sim 6$. These sources will contribute to $160\sim 254$ bursts per year.

Furthermore, observations suggest that inwards of S0-2's orbit ($\lsim 1000$~au) there is a hidden mass of $\sim 3000~\rm M_{\odot}$ \citep[e.g.,][]{Do+19,Gravity+20}. This is supported by theoretical arguments of the stability of S0-2, \citep[e.g.,][]{Naoz+20,Zhang+23,Will+23}.  
We, thus, explored the possibility that these unidentified objects are all stellar-mass BHs binaries formed within the recent $10^{7}~\rm yr$, and find that there can be $\sim 3- 20$ detectable RB sources with $\sim 10 - 300$ bursts per year under this assumption (see the grey data point in Figure~\ref{fig:population}). We note that the existence of BBHs in the inner $1000~\rm au$ is highly uncertain, but any potential population in this region can have a significantly high fraction of bursting systems. Therefore, GW burst detection in the future will serve as a strong constraint on the number of BBHs in the inner $1000~\rm au$ of the GN. For more details, see Appendix~\ref{append:GN}.

These results highlight the close relationship between the number of bursting BBHs and their formation history. In the young population of BBHs, RB systems will remain detectable up to $\sim 10^{7} ~\rm yr$ after being driven to extreme eccentricity by the SMBH, which is longer than the YNC's age. However, despite the large total mass of the old population, they are likely to reach a steady state because of the low formation rate of BBHs. Therefore the young population will appear to have a much higher RB source fraction than the average value, as depicted in Figure \ref{fig:number of source}. 

In other words, RB sources can serve as a tracer of active star formation and dynamical evolution of compact object binaries in the recent $\sim 10^{7}~\rm yr$ history of the central Milky Way. We speculate that excess detection of bursting sources in certain regions of the Milky Way may indicate a recent episode of compact object formation.

\subsection{EMRIs}
EMRIs occur when a stellar-mass BH merges with an SMBH, emitting GW in the process. One of the popular formation channels for EMRIs is based on weak gravitational interactions between neighboring objects in the dense environment surrounding SMBHs, known as two body relaxation/lose cone dynamics \citep[e.g.,][]{Merritt2010,Binney2008,Hopman_2007,Hopman2006,Alexander+09,Sari+19,Berry_2013,Fan22}. In Figure \ref{fig:population}, we depict the expected number of GW bursts from EMRIs (i.e., EMRBs) following \citet{Hopman_2007,Berry_2013, Fan22}, as well as the EMRBs rate assuming a significant fraction of SMBH binary existing in the universe \citep[see, e.g.,][]{Naoz+22,Naoz+23}.

In particular, \citet{Naoz+23} estimate that there can be as much as $100\sim 2000$ EMRIs detected per year if there is a significant fraction of galaxies holding an SMBH binary at their galactic center. Their simulation shows that most of the detectable EMRIs via SMBH binary channel are in a highly eccentric configuration, with the orbital period in the range of $\sim 1 - 10~\rm yr$. Therefore, we estimate the number of detectable bursting sources and bursts per year by multiplying the EMRIs number with their orbital frequency ($\sim 0.1 - 1~\rm yr^{-1}$). It turns out that under the assumption of \citet{Naoz+23}, there can be $\sim 1000 - 20000$ bursting EMRIs detected during the LISA mission, which contribute to $\sim 100 - 20000$ GW bursts per year. 

We emphasize that, similar to the case of bursting BBHs in the inner $1000~\rm au$ of the Milky Way (see Section~\ref{sec:GN}), the existence of SMBH binaries in the universe is poorly constrained. Therefore, in Figure~\ref{fig:population}, we use a hollow diamond to highlight the uncertainty of EMRI bursting sources formed via the SMBH binary channel. Similar to the arguments in \citet{Naoz+23}, whether or not we will detect a large number of bursting sources as expected by this channel can strongly constrain the fraction of SMBH binaries in the universe.

\section{Discussion}
\label{sec:discussion}
Many dynamically formed GW sources are expected to undergo an evolutionary stage at which the compact object binary is driven to extreme eccentricity. Such a system will emit GW bursts effectively upon each pericenter passage (see, for example, Figure \ref{fig:diffe}), creating a pulse-like pattern in the signal before finally losing enough orbital energy and becoming a merger. Therefore, GW bursts from highly eccentric compact object binaries can be a promising tracer of their dynamical formation. 

In Section~\ref{sec:RB property}, we estimate the detectability and lifetime of these sources analytically  (see Eqs.~(\ref{eq:lifetime})-(\ref{eq:SNR2})) and compare the results to numerical calculations (see Figure~\ref{fig:maps}). Particularly, we show that stellar-mass bursting sources should be detectable in the Milky Way ($\sim \rm 10 kpc$), with a much longer lifetime than other mHz sources (up to $10^{7}$~yr).
Moreover, a bursting source yields a larger strain amplitude than a low eccentricity source with the same average signal-to-noise ratio (see Figure \ref{fig:LISA2systems}). Considering this feature, we can potentially enhance these sources' detectability and parameter extraction in future data analysis.

Notably, BBH systems in many different regimes, such as in the field, globular clusters, and galactic nuclei, can naturally form bursting GW sources in the mHz band. For example, in Section~\ref{sec:population}, we show that bursting sources can dominate the population of mHz BBHs in the Milky Way since their longer lifetime (Equation~(\ref{eq:RBtime})) results in a larger number expectation (Equation~(\ref{eq:steadystate})). Adopting the constraints from LIGO's observation, we estimate $\sim 10$ bursting BBHs detectable for the future LISA mission under the steady-state approximation (see Section~\ref{sec:ligoconstrain}). 

In Figure \ref{fig:population}, we present the estimated number of bursts as a function of bursting sources for different formation environments. In particular, we consider stellar-mass BHs in the galactic field, the globular clusters, and the galactic nucleus. We find that these channels can contribute to the number of detectable stellar-mass bursting BBHs in a range of $3\sim 45$, with $10^{2}\sim 10^{4}$ mHz GW bursts observed during the LISA mission.

We highlight that the number of detectable bursting sources can well exceed the steady-state value since the formation history of bursting sources significantly affects their number in the mHz band. In particular, the Milky Way is expected to be a star-forming galaxy. Therefore, in Section~\ref{sec:GN}, we go beyond the steady-state approximations and simulate the time evolution of bursting sources in the galactic center Young Nuclear Star Cluster, which is expected to form $\sim2 - 8 ~\rm Myr$ ago. The simulation result supports that long-living bursting sources can remain detectable for $\sim 10^{7}~\rm yr$. This region, with an active star formation in the recent few million years, will have a much higher fraction of bursting sources ($2\sim 4$ bursting sources out of $\sim 100$ BBHs) than those old ones ($\sim 1$ bursting sources out of $\sim 1500$ BBHs). In future observation, the distribution of bursting sources, as well as their actual number, can serve as a valuable tool to probe the GW sources' creation time in different regions of the Milky Way. 

To conclude, unlike other mHz sources with moderate eccentricity, the detection of stellar-mass bursting systems is mostly limited in the Milky Way and nearby galaxies (see, e.g., Figure~\ref{fig:maps2}). However, these sources' bursting nature also results in a long lifetime and a potentially large number in the entire mHz population. Therefore, it is very likely that we can find many bursting sources at a close distance (e.g., a few dozen bursting BBHs in the Milky Way), with hundreds to thousands of GW bursts detected in the LISA observation. By studying the properties of these sources, we can better understand the early stages of dynamical formation and map the compact objects' formation history in our galaxy.

\acknowledgments
The authors thank Cheyanne Shariat, Sanaea Rose, Brenna Mockler, Xian Chen, and Xiao Fang for their valuable discussions and the anonymous referee for their useful comments. ZX acknowledges partial support from the Bhaumik Institute for Theoretical Physics summer fellowship. ZX, SN, and EM acknowledge the partial support from NASA ATP 80NSSC20K0505 and from NSF-AST 2206428 grant and thank Howard and Astrid Preston for their generous support. This work was supported by the Science and Technology Facilities Council Grant Number ST/W000903/1 (to BK).

\appendix
\begin{figure}[htbp]
    \centering
    \includegraphics[width=3.2in]{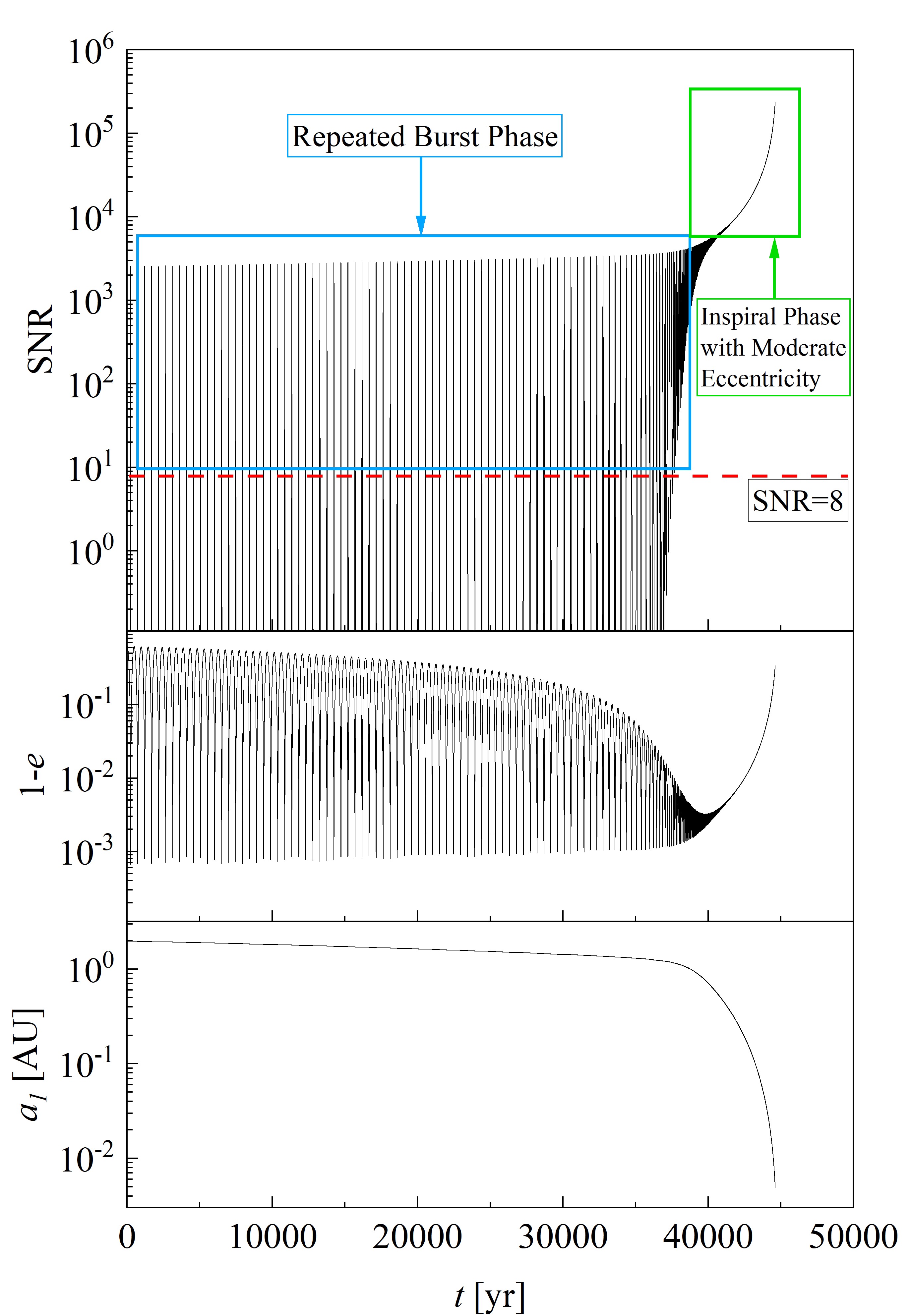}
    \caption{{\bf{An example of the repeated burst phase and inspiral phase during the evolution of an EKL merger.}}
    We show a BBH system with  $m_{1}=57.7~\rm {M_{\odot}}, m_{2}=51.9~\rm {M_{\odot}},$ 
    $a_{1}=1.97 ~\rm au,$ 
    $e_{1}=0.38, $
    placed near a SMBH with $\rm{M} = 4\times 10^{6}~\rm M_{\odot}, $
    $a_{2}=4706 ~\rm au,$
    $e_{2}=0.97,$
    $i=88^{\circ},$
    $D_{l}=8~\rm kpc$
    . {\it Upper Panel} shows the signal-to-noise ratio of this source, for the observational time $4~\rm yr$. 
    {\it Middle Panel} and {\it Bottom Panel} show its eccentricity and semi-major axis evolution as functions of time.
    As is shown in the figure, this example system undergoes strong EKL oscillation because of the SMBH. It spends a significant fraction of time in the repeated burst phase, bursting detectable GW signal via highly eccentric orbit, before finally reaching the inspiral phase with moderate eccentricity. 
    }
    \label{fig:EKLevolution}
\end{figure}

\section{details of numerical simulations}
\label{append:details}
\subsection{BBHs in the Field}
\label{append:field}
For the BBHs in the galactic field, flyby gravitational interactions with other neighbors may excite their high eccentricity, driving the binary into a merger \citep[e.g.,][]{Michaely+19,Michaely+22}. In this work, we adopt the model in \citet{Michaely+22} to calculate the properties of bursting BBHs born from this channel. 

In particular, we choose the configuration of a Milky-Way type galaxy, with the stellar density profile and wide BBHs fraction following Eq.~(23)$\sim$(26) in \citet{Michaely+19}. For simplicity, we further assume that all the BBHs have the mass of $10-10~\rm M_{\odot}$, with the log-uniform distribution in the semi-major axis (from $100$ to $50000~\rm au$). Second, we run a Monte-Carlo simulation, randomly choose the position and semi-major axis of BBHs following the distribution mentioned above, and calculate the fly-by merger rate as a function of semi-major axis \citep[see Eq.~(16)$\sim$(23) in][]{Michaely+22}.

After undergoing a fly-by interaction, the BBHs with a given semi-major axis need to have the eccentricity exceed a threshold, $e_{\rm crit}$, to make themselves merge before the next fly-by \citep[see Eq.~(2) in][]{Michaely+22}. Therefore, we randomly generate the eccentricity of fly-by induced merger in the range of $e_{\rm crit}$ to $1$, following the thermal distribution $F(e)=2e$. 

Using the methodology mentioned above, we can specify each merger system's semi-major axis and eccentricity in the simulation. This information allows us to calculate how many bursts they emit, and how long they stay detectable. For example, we can evolve a system following Eq.~(\ref{eq:isolated ae}) and trace its signal-to-noise ratio using Eq.~(\ref{eq:SNR2}). Once the system becomes detectable, the merger timescale and number of bursts can be calculated via integration (see the example track in Figure.~\ref{fig:maps} for demonstration). The number of bursting systems in the fly-by channel is calculated using the steady-state approximation. In other words, we multiply the corresponding merger rate with the average detectable time for each semi-major axis, then do the summation over different semi-major axis to get the expectation of sources' number (see Eq.~(\ref{eq:steadystate})). 

Assuming a $10$~yr LISA observation with the signal-to-noise ratio threshold $5$, we got the expectation of fly-by induced, bursting field BBHs number is $N_{\rm field}\sim 3.3$ in the mHz band, with the number of detectable bursts $N_{\rm burst, field}\sim 61$ per year and merger rate $\Gamma_{\rm field}\sim 5\times 10^{-7} \rm yr^{-1}$ in the Milky Way (or $\sim 10 ~\rm Gpc^{-3}yr^{-1}$ in the universe). Among these sources, we identified $\sim 1.8$ repeated burst source and $\sim 1.5$ non-repeated burst source (i.e., $T_{\rm orb}>10~\rm yr$).  

According to \citet{Michaely+19}, the BBH merger rate from the field channel can be within the range of $5_{-3}^{+5} \mathrm{Gpc}^{-3} \mathrm{yr}^{-1}$ in the local universe. Moreover, the number of bursts and bursting sources is proportional to the merger rate of BBHs. Therefore, to get a realistic estimation of the bursting sources' properties, the numbers we got from the simulation should be multiplied by a factor of $0.5_{-0.2}^{+0.5}$, which is reflected in the results in Section~\ref{sec:field}.

\subsection{BBHs in Globular Clusters}
\label{append:gcs}
As is shown in Section~\ref{sec:gcs}, we use the eccentricity distribution of BBHs at a given frequency \citep[see Fig.~4 in ][]{Martinez+20} and the spatial distribution of globular clusters in the Milky Way \citep{Arakelyan_2018} to generate the initial condition of bursting BBHs after undergoing dynamical interaction in GCs. For conservation purposes, we exclude the single-single capture and few body capture channels since the GW sources from these two channels are mostly formed above the mHz band, thus may not be suitable targets for LISA \citep{Kocsis2020}.

Adopting the same method as is shown in Section~\ref{append:field}, we carried out Monte-Carlo Simulations and generated the parameters of bursting BBHs following the distribution mentioned above \citep{samsing18,Orazio+18,Martinez+20}, then evolved them to the merger. The number of bursting sources and bursts is calculated under the steady-state approximation, taking into account the burst sources' properties from different channels of merger in GCs. Particularly, for a given sub-channel in GCs, we get the averaged bursting lifetime and the total number of bursts from the simulation, then multiply it with the merger rate (see Eq.~(\ref{eq:steadystate})). 

Assuming a $10$~yr LISA observation with the signal-to-noise ratio threshold $5$ and GCs merger rate $\Gamma_{\rm GC}\sim 1\times 10^{-6} ~\rm yr^{-1}$ in the Milky Way ( $\sim 20 ~\rm Gpc^{-3}yr^{-1}$ in the universe), we find the number of detectable bursting BBHs in the GCs of Milky Way to be $N_{\rm GC}\sim 14$, with the number of detectable bursts $N_{\rm burst, GC}\sim 176$ per year. Among these sources, we identified $\sim 4$ repeated burst source and $\sim 10$ non-repeated burst source. Similar to the case of field binaries in Section~\ref{sec:field}, the number of bursting sources is proportional to the merger rate ($\mathcal{R}_0=7.2_{-5.5}^{+21.5} \mathrm{Gpc}^{-3} \mathrm{yr}^{-1}$ \citep{Rodriguez2016,Kremer_2020,Antonini_2020}). Therefore, the result in the simulation should be multiplied by a factor of $0.36_{-0.26}^{+1.12}$, which is reflected in the results of Section~\ref{sec:gcs}.

\subsection{BBHs in the Galactic Nucleus}
\label{append:GN}
Stellar-mass BBHs surrounding the supermassive black hole in the galactic nucleus will naturally be in the configuration of a hierarchical triple system, thus undergoing eccentricity oscillation via the EKL mechanism. Based on the observational results, we carried out Monte-Carlo simulations of these BBHs' evolution. The simulations of hierarchical triple systems include the secular equations up to the octupole level of approximation \citep{Naoz+13}, general relativity precession \citep{naoz13}, and GW
emission \citep{Peters64}. In particular, we take into account three different populations of BBHs in the galactic nucleus:
\begin{enumerate}
    \item {\bf BBHs from the Main Population of Stars in the Galactic Nucleus.} This population is expected to have a distance to SMBH within $5~\rm pc$, age $\sim 2- 8~\rm Gyr$, total mass $\sim 1.8\times 10^{7}$~M$_{\odot}$ \citep[see, e.g.,][]{Pfuhl11,Launhardt02}. In particular, we randomly generate BBHs with log uniform distribution in mass and semi-major axis, ranging from $6$ to $100$~M$_{\odot}$ and $0.1$ to $50$~au, respectively. For the spatial distribution of these BBHs, we adopt the isotropic distribution, with the radial density profile in the galactic center following \citet{Hoang+18}. The number of systems is calibrated using the $m-\sigma$ relation. For conservation purposes, we rule out the systems that are too close to the SMBH to be classified as hierarchical triple systems \citep{Naoz16}. 

    After generating the initial condition for BBHs, we evolve these systems numerically, counting their bursting time and number of emitted bursts. The integration is stopped once the system becomes a GW merger or reaches the evaporation timescale \citep[see Eq.(3) in][]{Hoang+18}. Because of the old age of the main population, we use the steady-state approximation to work out the expectation of bursting sources' number, i.e.:
    \begin{equation}
        N_{\rm RB}=\Gamma_{\rm rep, all} f_{\rm RB} \tau_{\rm RB},
    \end{equation}
    in which $\Gamma_{\rm rep, all}$ is the replenishing rate of BBHs in the GN, $f_{\rm RB}$ is the fraction of BBHs in the GN that become a repeated burst source, and $\tau_{\rm RB}$ is the average lifetime time of bursting sources when they are detectable. Under the steady-state approximation, the total replenishing rate, $\Gamma_{\rm rep, all}$, equals the total empty rate at which BBHs merge or evaporate. Therefore, in the simulation, we determine the $\Gamma_{\rm rep, all}$ by multiplying the inverse of all the BBHs' average lifetime with their total number ($\sim 1500$) in the inner $0.5\, \rm pc$. 
    
    From the result of the simulation, $\Gamma_{\rm rep, all}\sim 3\times 10^{-6}~\rm yr^{-1}$. We note that this quantity equals the combined rate of GW merger plus evaporation, thus is higher than the GW merger rate constrained by LIGO. Moreover, the fraction of BBHs that turns into a bursting source is $f_{\rm RB}\sim 3\%$, and the average bursting time is $\tau_{\rm RB}\sim 1.4\times 10^{7}~\rm yr$. Therefore, we estimate the number of main population bursting sources as $\sim 1$ in the inner $0.5$~pc of the Milky Way. A similar approach can be applied to calculate the number of bursts, and the result from the simulation is $\sim 100$~$\rm yr^{-1}$.
    
    \item {\bf BBHs in the Young Nuclear Star Cluster} According to the observation \citep{Paumard06,Lu2009,Lu2013}, the YNC has a distance to SMBH within $\sim 0.5~\rm pc$, age $\sim 2- 8~\rm Myr$, and total mass $\sim 1.4 - 3.7\times 10^{4}M_{\odot}$, with top-heavy mass distribution. Because of their small mass and young age, we need to go beyond the steady-state approximation and simulate the time evolution of bursting GW sources as a function of time. 
    
    For simplicity, we assume that this YNC is born in a starburst at $t=0$, and all the massive stars are in the binary system. After evolving for a few $\rm Myr$, the massive stars have all died and become black holes, and we get the corresponding blackhole mass following the results in \citet{woosley02,Belczynski_2020}. According to the simulation result, there will be $\sim 100 - 400$ BBHs newly formed in the YNC. Because of the existence of mass gap \citep[see, e.g.,][]{Bond84,Fryer01}, most of these black holes have mass $\sim 10$~M$_{\odot}$. We distribute the BBHs isotropically around the SMBH, with the same density profile as the main population, and trace their evolution as hierarchical triple systems.
    
    As is discussed in Section~\ref{sec:GN}, although the YNC has a relatively small total mass, its bursting BBHs number can be as much as 4, which is much higher than the main population with old age and a slow replenishing rate. We show the simulation results in Figure.~\ref{fig:number of source}. For completeness, we also show the expected number of bursts detected during the LISA mission as a function of the YNC's age (see Figure~\ref{fig:appendGN}).
    \begin{figure}[htbp]
    \centering
    \includegraphics[width=3in]{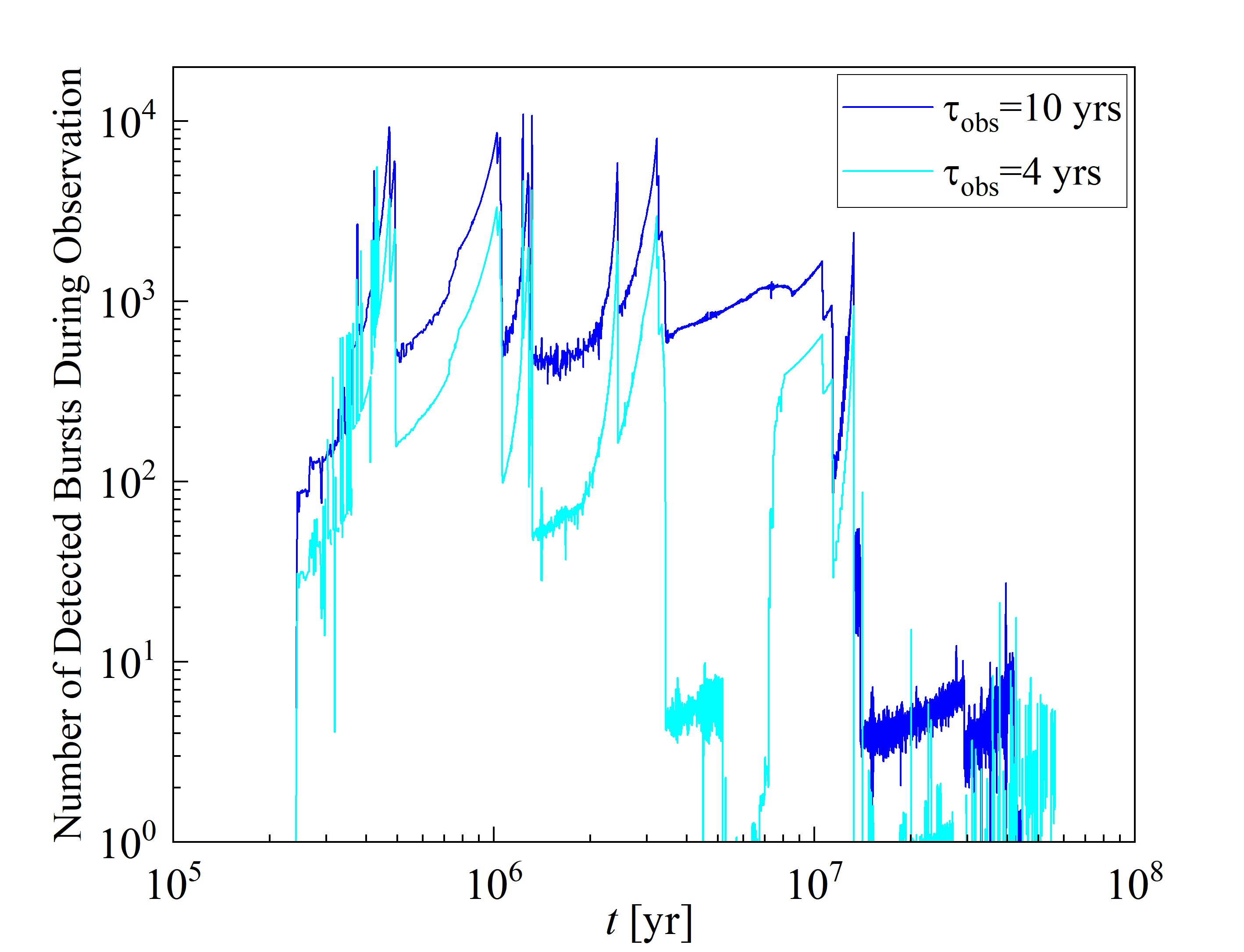}
    \caption{{\bf{Number of detectable bursts during LISA mission, as a function of the YNC's age.}}
    Here we show the simulation result of observable bursts in the Milky Way center young nuclear cluster, as a function of the cluster's age. The deep blue line assumes the LISA observational period $\tau_{\rm obs}= 10 ~\rm yr$ while the light blue line represents $\tau_{\rm obs}= 4 ~\rm yr$. The large fluctuation of the light blue line is caused by the limited number of systems in the simulation.
    }
    \label{fig:appendGN}
\end{figure}
    \item {\bf BBHs in the Inner $1000$~au of the Galactic Nucleus} The observation of stellar dynamics in the galactic center suggests that there can be up to $\sim 3000$~M$_{\odot}$ unknown mass within the inner $1000$~au of the SMBH \citep[see, e.g.,][]{Ghez+08,Gravity+20,Will+23}. Therefore, we explored the possibility that these unknown objects are made up of stellar-mass BBHs (i.e., $\sim 150$ BBHs). Since we are only aiming at a heuristic estimation, the simulation assumes that all these BBHs have the mass of $10-10~\rm M_{\odot}$, with the spatial density profile the same as for bursting BBHs in the YNC, and was born at $t=0$.

    Adopting the same approach as modeling the time evolution of bursting sources in the YNC, we get the properties of this inner $1000$~au population (see the hollow grey triangle in Figure~\ref{fig:population}). In particular, the existence of BBHs within the inner $1000$~au, if any, will give a significantly higher fraction of bursting sources.  When the cluster is a few Myr old, up to $\sim 13\%$ of all the BBHs can emit GW bursts, contributing to $\sim 20$ bursting sources and $200$ bursts per year. On the other hand, if there is no such bursting source in future observation, we can put stringent constraints on the existence of BBHs in the center of our galaxy.
\end{enumerate}

\section{Numerical Approach to calculate the SNR of highly eccentric binaries}
\label{append:snr}
The time domain waveform of eccentric binaries, $h(a, e, t)$ , can be decomposed into different harmonics, with the frequency $f_n=nf_{\rm orb}$ \citep{peters63,Kocsis_2012}:
\begin{equation}
h(a, e, t)=\sum_{n=1}^{\infty} h_n\left(a, e, f_n\right) \exp \left(2 \pi i f_n t\right),
\end{equation}
where
\begin{equation}
h_n\left(a, e, f_n\right)=\frac{2}{n} \sqrt{g(n, e)} h_0(a),
\end{equation}
and:
\begin{equation}
h_0(a)=\sqrt{\frac{32}{5}} \frac{m_1 m_2}{D_l a}
\end{equation}

\begin{eqnarray}
g(n, e) & =\frac{n^4}{32}\left[\left(J_{n-2}-2 e J_{n-1}+\frac{2}{n} J_n+2 e J_{n+1}-J_{n+2}\right)^2\right. \\
& \left.+\left(1-e^2\right)\left(J_{n-2}-2 J_n+J_{n+2}\right)^2+\frac{4}{3 n^2} J_n^2\right],
\end{eqnarray}
in which $J_i$
is the i-th Bessel function evaluated at $ne$. 

The integral of signal-to-noise ratio, in the frequency domain, is defined as \citep[see, e.g.,][]{Robson+19,chen19}:
\begin{equation}
{\rm{SNR}}^2(a, e)=\int \frac{h_c^2(a, e, f)}{f^2 S_n(f)} d f,
\end{equation}
in which $h_c$ is the characteristic strain:
\begin{equation}\label{eq:hc}
h_c^2(a, e, f)=4 f^2|\tilde{h}(a, e, f)|^2,
\end{equation}
and $\tilde{h}(a, e, f)$ is the Fourier transform of the GW signal \citep{moore15}.

For highly eccentric binaries, the transient nature makes its frequency domain waveform split into millions of harmonics (peaks, see Figure~\ref{fig:LISA2systems}). They are separated by an interval of $f_{\rm orb}$, and each peak has the width of $\Delta f\sim n\dot{f}_{\rm orb} T_{\rm obs}$ (caused by the orbital frequency shift of GW sources). Therefore, we can transform the expression of SNR into the summation of harmonics \citep[see appendix A of ][for details]{Xuan_2026}:
\begin{equation}
    {\rm SNR}^2=8h^2_0(a)\sum_n\frac{g(n,e)}{S_n(nf_{\rm orb})n^2}T_{\rm obs} \ .
    \label{eq:snrsum}
\end{equation}
As is shown in Equation~(\ref{eq:snrsum}), the huge number (millions) of harmonics adds up to the numerical difficulty of calculating the SNR. However, we can further simplify this expression using the knowledge of the envelope of the frequency spectrum. In other words, instead of directly summing over millions of peaks, we can calculate the averaged amplitude in a wider frequency bin, and work out the density of peaks below that envelope (area enclosed by the power spectrum), then get a useful approximation of the SNR.

In particular, let's consider the integration in the frequency domain, take a value of $f$ and the integration interval $df$ around it. For each integration interval, the number of peaks contained in this frequency bin is $\Delta n\sim df/f_{\rm orb}$, while the amplitude of these peaks can be represented by the average value $g(n_{\rm avg},e)$, with $n_{\rm avg}=f/f_{\rm orb}$.

Thus, the summation in Equation~(\ref{eq:snrsum}) can be turned into the integration of a smoothed function:
\begin{eqnarray}
    {\rm SNR}^2=8h^2_0(a)\int\frac{g(n_{\rm avg},e)}{S_n(f)(\frac{f}{f_{\rm orb}})^2}T_{\rm obs}\frac{df}{f_{\rm orb}}
    =8h^2_0(a)f_{\rm orb}T_{\rm obs} \int\frac{g(n_{\rm avg},e)}{S_n(f)f^2}df 
    \label{eq:snrsum2}
\end{eqnarray}
Equation~(\ref{eq:snrsum2}) serves as a fast numerical estimation of highly eccentric binaries' SNR. The accuracy of integration can be changed by adjusting $d f$. In the limit of $d f=f_{\rm orb}$, we recover the strict expression of  Equation~(\ref{eq:snrsum}).


\bibliography{bibbase}
\end{document}